\documentclass[aps,pre]{revtex4}
\usepackage[dvips]{graphics}
\usepackage{graphicx}
\usepackage{amsfonts}
\usepackage{amssymb}
\usepackage{amsmath}
\usepackage{subfigure}
\usepackage{color}

\begin{document}

\title{Effects of Long-Range Nonlinear Interactions in Double-Well Potentials%
}
\author{C. Wang$^{1}$, P. G.\ Kevrekidis$^{1}$, D. J.\ Frantzeskakis$^{2}$,
and B. A. Malomed$^{3}$ }
\affiliation{$^{1}$ Department of Mathematics and Statistics, University of
Massachusetts, Amherst MA 01003-4515, USA \\
$^{2}$ Department of Physics, University of Athens, Panepistimiopolis,
Zografos, Athens 15784, Greece \\
$^{3}$ Department of Physical Electronics,\ School of Electrical
Engineering, Faculty of Engineering, Tel Aviv University, Tel Aviv 69978,
Israel}

\begin{abstract}
We consider the interplay of linear double-well-potential (DWP) structures
and nonlinear long-range interactions of different types, motivated by
applications to nonlinear optics and matter waves. We find that, while the
basic spontaneous-symmetry-breaking (SSB) bifurcation structure in the DWP
persists in the presence of the long-range interactions, the critical points
at which the SSB emerges are sensitive to the range of the nonlocal
interaction. We quantify the dynamics by developing a few-mode approximation
corresponding to the DWP structure, and analyze the resulting system of
ordinary differential equations and its bifurcations
in detail. We compare results of this
analysis with those produced by the full partial differential equation,
finding good agreement between the two approaches. Effects of the
competition between the local self-attraction and nonlocal repulsion on the
SSB are studied too. A far more complex bifurcation structure involving
the possibility for not only supercritical but also subcritical
bifurcations and even bifurcation loops is identified in that case.
\end{abstract}

\date{\today}
\maketitle

\section{Introduction}

The studies of Bose-Einstein condensates (BECs) \cite{book1,book2,ourbook}
and nonlinear optics \cite{agra} keep drawing a great deal of attention due
to experimental advances in versatile realizations of such systems, as well
as considerable progress in the analysis of relevant models 
based on the nonlinear Schr{\"{o}}dinger (NLS) -type equations. One of remarkable
features specific to these fields is the possibility of tailoring particular
configurations by dint of suitably designed magnetic and/or optical trapping 
mechanisms (possibly acting in a combination) that confine the atoms in the
case of BEC, or virtual (photonic) and material structures manipulating the
transmission of light in nonlinear optical media. These achievements 
motivate the detailed examination of the existence, stability and dynamical
behavior of nonlinear modes in the form of matter or optical waves. The NLS
equation \cite{agra,sulem}, as well as its 
variant known as the
Gross-Pitaevskii (GP) equation \cite{book1,book2,ourbook} in the BEC context 
are often at the center of such analysis.

Within the diverse range of external confinement mechanisms, one that has
attracted particular attention is that provided by double-well potentials
(DWPs). Its prototypical realization in the context of BEC relies on 
the combination of a parabolic (harmonic) trap with a periodic potential,
which can be created, as an ``optical lattice", by the interference of laser
beams illuminating the condensate \cite{Morsch}. The use of a DWP created as
a trap for BEC (with the intrinsic self-repulsive nonlinearity) has revealed
a wealth of new phenomena in recent experiments \cite{markus1}, including
the tunneling and Josephson oscillations for small numbers of atoms in the
condensate, and macroscopic quantum self-trapped states for large atom
numbers. Prior to this work, as well as afterwards, motivated by its
findings, a wide range of theoretical studies investigated such DWP
settings, including such issues as finite-mode reductions and
symmetry-breaking bifurcations \cite{smerzi,kiv2,mahmud,bam,Bergeman_2mode,infeld,todd,theo}, quantum effects \cite{carr}, and nonlinear variants of the DWP \cite{pseudo}. DWP settings
and spontaneous-symmetry-breaking (SSB) effects in them have also been
studied in nonlinear-optical settings, such as formation of asymmetric
states in dual-core fibers \cite{fibers}, self-guided laser beams in Kerr
media \cite{HaeltermannPRL02}, and optically-induced dual-core waveguiding
structures in photorefractive crystals \cite{zhigang}.

One of recent developments in 
both fields of 
matter and optical waves is 
the study of effects of long-range nonlinear interactions. In the BEC these
studies are dealing with condensates formed by magnetically polarized $^{52}$%
Cr atoms \cite{Cr} (see recent review \cite{review}), dipolar molecules \cite%
{hetmol}, or atoms in which electric moments are induced by a strong
external field \cite{dc}. 
Matter-wave solitons supported by the
dipole-dipole interactions were predicted in isotropic \cite{Pedri05},
anisotropic \cite{Tikhonenkov08}, and discrete \cite{2Ddiscrete}
two-dimensional (2D) settings, and in the quasi-1D configurations \cite%
{Sinha,us} (the latter was done not only in BEC, but also in a model of the
Tonks-Girardeau gas \cite{BBB}). In optics, prominent examples of patterns
supported by long-range effects are stable vortex rings predicted in media
with the thermal nonlocal nonlinearity \cite{Krolik}, as well as the
experimental realization of elliptically shaped spatial solitons in these
media \cite{moti1}.

Our aim in the present work is to examine effects of long-range interactions
in the context of DWPs. This is a topic of increasing current interest; in
the context of dipolar multi-dimensional condensates, it was recently
addressed in Refs.~\cite{pra09}, where the phase diagram of the system was
explored, as a function of the strength of the barrier in the DWP, number of
atoms, and aspect ratio of the system. The possibility of a transition from
a symmetric state to an asymmetric one, and finally to an unstable
higher-dimensional configuration was considered. Here, we focus on the 1D
setting, and explore different types of long-range interactions, including
the dipole-dipole interactions, as motivated by Refs.~\cite{Sinha,us,BBB}, 
as well as the interactions with Gaussian and exponential
kernels, motivated by nonlinear-optical models \cite{krol1}. In particular,
we consider the effect of the range of the interaction, with the objective
to consider a transition from the contact interactions to progressively
longer-range ones. We conclude that the phenomenology of the short-range
interactions persists, i.e., the earlier discovered SSB phenomena \cite%
{Bergeman_2mode,todd,theo,shliz} still arise in the present context.
However, the critical point of the SSB transitions features a definite,
monotonically increasing, dependence on the interaction range, which is
considered in a systematic way. More elaborate scenarios can be detected in
the case where in addition to the long-range interactions, there is a
competing short-range component. 
In such a case, we identify not only the earlier
symmetry-breaking bifurcations but also reverse, ``symmetry-restoring''
bifurcations, as well as the potential for symmetry-breaking to arise (for
the same parameters) both from the symmetric and from the antisymmetric
solution branch. Both of these are phenomena that, to the best of our
knowledge, have not been reported previously.  Interestingly, the only
example where subcritical bifurcations have been previously
discussed in the double well setting for GP equations 
is the very recent one of extremely (and hence somewhat unphysically) 
high nonlinearity exponents in \cite{sacchetti}.

The presentation is structured as follows. In section II, we present the
model and the quasi-analytical two-mode approximation, which clearly reveals
the system's bifurcation properties. In section III, we corroborate these
analytical predictions by full numerical results for different types of the
long-range kernel. In section IV, we consider a more general case, in which
the long-range nonlinear repulsion competes with the local attraction. There,
we illustrate how the
competition strongly affects the character of the SSB in the DWP setting. 
Finally, in
section V, summarize our findings and present our conclusions.

\section{The long-range model and the analytical approach}

\label{secAnaly}

In the quasi-1D setting, the normalized mean-field wave function $\psi (x,t)$
obeys the scaled GP equation,
\begin{equation}
i\partial _{t}\psi +\mu \psi ={\cal L}\psi +s\left[ \int_{-\infty }^{\infty
}K\,(x-x^{\prime })|\,\psi (x^{\prime })|^{2}\,dx^{\prime }\right] \,\,\psi ,
\label{eq1}
\end{equation}%
where $\mu $ is the chemical potential, and
\begin{equation}
{\cal L}=-(1/2)\partial _{x}^{2}+V(x)  \label{eq2}
\end{equation}%
is the usual single-particle energy operator, which includes the confining
DWP
\begin{equation}
V(x)=(1/2)\hat{\Omega}^{2}x^{2}+V_{0}\,\mathrm{sech}^{2}\left( x/W\right) ,
\label{eq3}
\end{equation}%
with $\hat{\Omega}$ the normalized harmonic-trap's strength;
$\hat{\Omega} \ll 1$ in a quasi-1d situation in BECs. The nonlinear
term with coefficient $s$ accounts for the long-range interatomic
interactions, $s=\pm 1$ corresponding to the repulsion and attraction,
respectively. 
Note that the contact interaction is not taken into regard in Eq.~(\ref{eq1}), as we
aim to focus on the effect produced by the long-range nonlinearity (in the gas
of $^{52}$Cr atoms, the contact interaction may be readily suppressed by
means of the Feshbach resonance \cite{Cr}). In this work, we consider mainly
symmetric spatial kernels in Eq.~(\ref{eq1}), that are positive definite,
with the following three natural forms chosen for detailed analysis: the
Gaussian,
\begin{equation}
K(x)=\frac{1}{\sigma \sqrt{\pi }}\mathrm{exp}\left( -\frac{x^{2}}{\sigma ^{2}%
}\right) ,  \label{eq4}
\end{equation}
the exponential,
\begin{equation}
K(x)=\frac{1}{2\sigma }\mathrm{exp}\left( -\frac{|x|}{\sigma }\right) ,
\label{eq5}
\end{equation}%
and the \textit{cut-off (CO)} 
(alias generalized Lorentzian) kernel,
\begin{equation}
K(x)=\frac{10}{\pi }\sigma ^{3}(x^{2}+\sigma ^{2})^{-3/2}.  \label{eq6}
\end{equation}%
The width of the kernels, $\sigma $, determines the degree of the
nonlocality. All three kinds of the kernels go over into the $\delta $%
-function as $\sigma $ approaches zero, in which case Eq.~(\ref{eq1}) turns
into the usual local NLS/GP equation. All the kernels are normalized as the $%
\delta $-function, i.e., $\int_{-\infty }^{+\infty }K(x)dx=1$, and the norm
of the wave function will be used in the usual form, $N=\int_{-\infty
}^{+\infty }|\psi (x,t)|^{2}dx$. In what follows below, we adopt typical
physically relevant values of the scaled parameters, namely, $\hat{\Omega}%
=0.1$, $V_{0}=1$ and $W=0.5$, in which case the two lowest eigenvalues of
linear operator ${\cal L}$ with potential (\ref{eq3}) are numerically found to be $%
\omega _{0}=0.1328$ and $\omega _{1}=0.1557$.

\subsection{The two-mode approximation}

\begin{figure}[tbph]
\centering
\includegraphics[width=.3\textwidth]{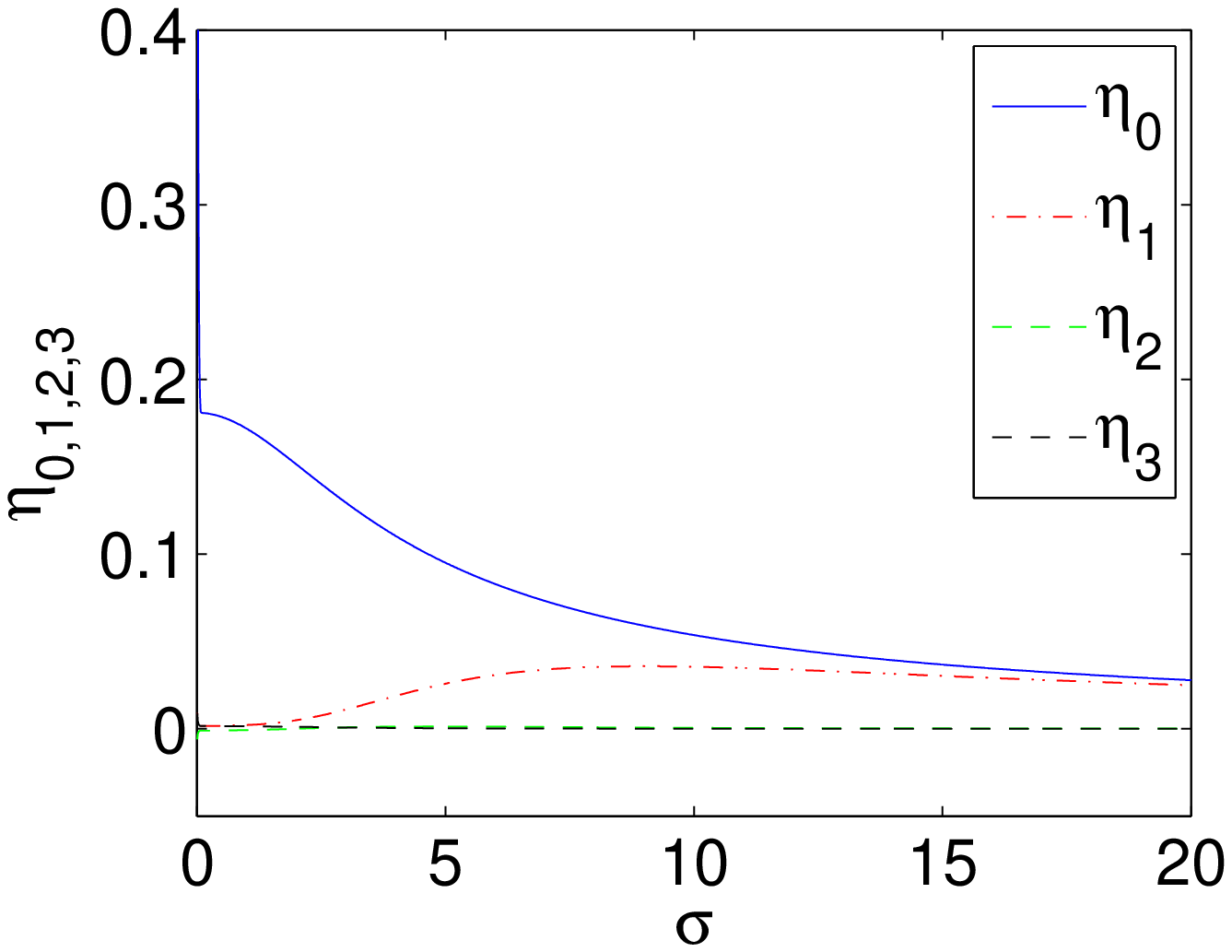} %
\includegraphics[width=.3\textwidth]{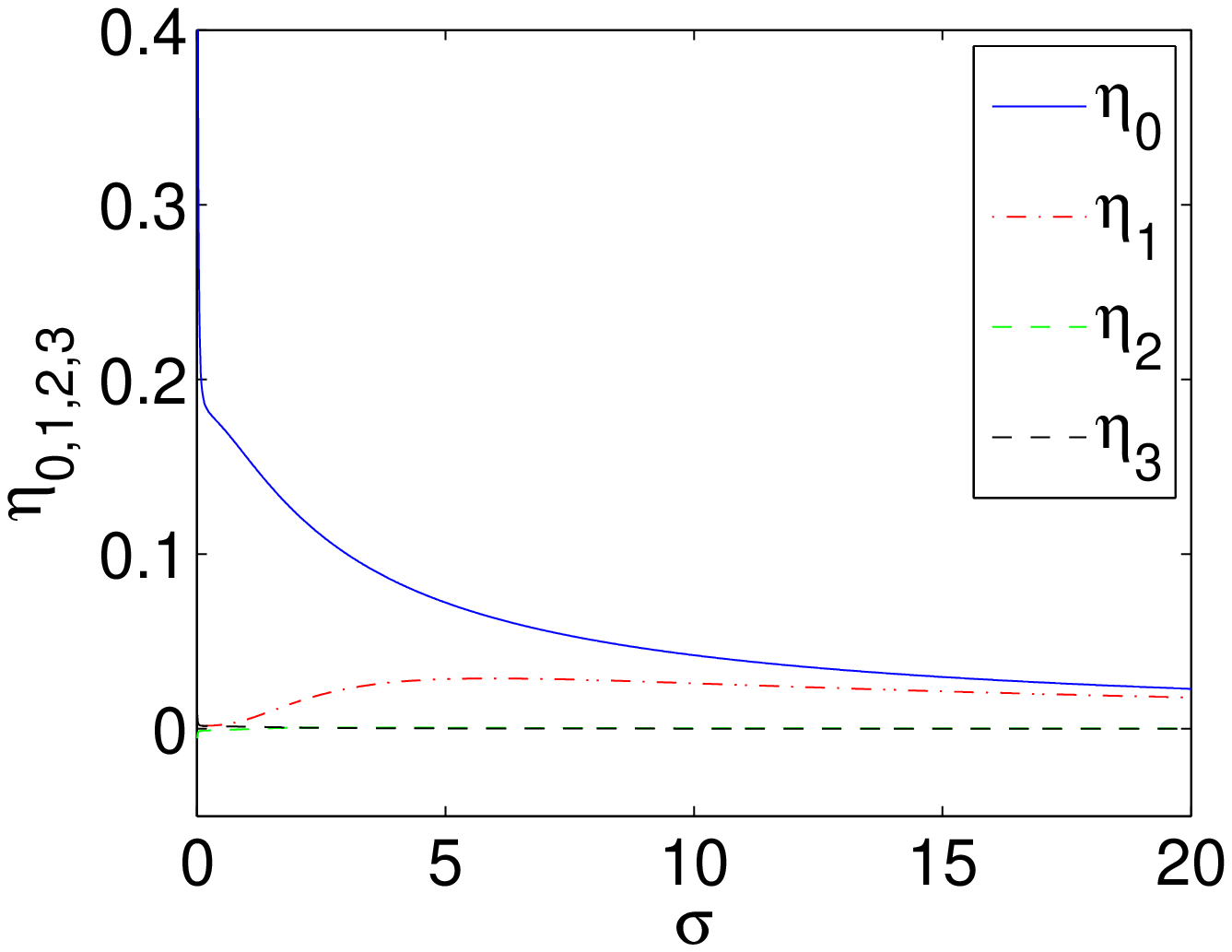} %
\includegraphics[width=.3\textwidth]{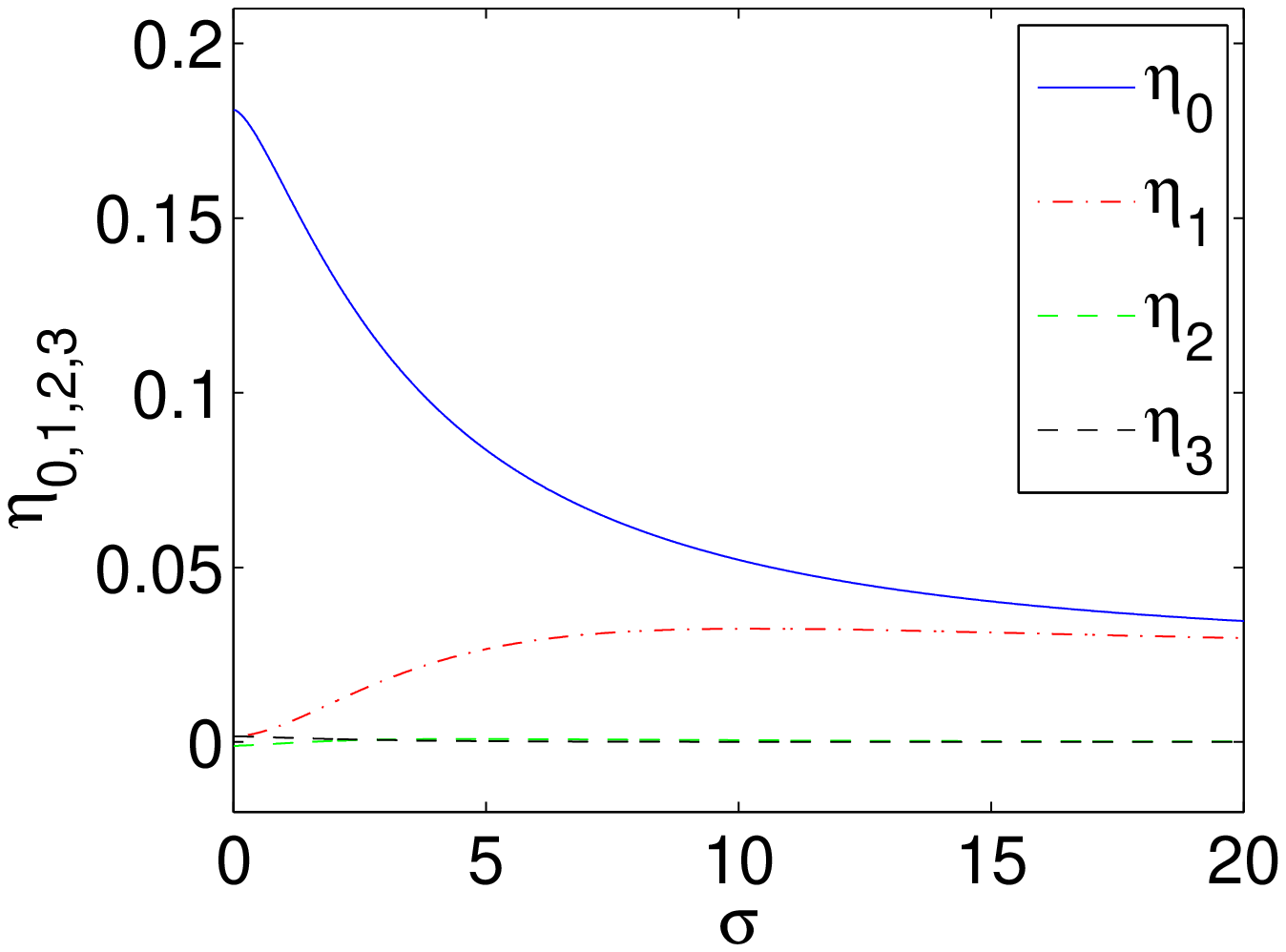}\newline
\caption{(Color online) The overlap integrals $\protect\eta _{0}$, $\protect%
\eta _{1}$, $\protect\eta _{2}$ and $\protect\eta _{3}$ as functions of
width $\protect\sigma $ of the kernels of the following types: Gaussian
(left), exponential (middle), and Lorentzian (right), as defined in Eqs. (%
\protect\ref{eq4})-(\protect\ref{eq6}).}
\label{fig1}
\end{figure}

The spectrum of the underlying linear Schr{\"{o}}dinger equation ($s=0$)
consists of the ground state, with wave function $u_{0}(x)$, and excited
states, $u_{l}(x)$ ($l\geq 1$). In the weakly nonlinear regime, the
Galerkin-type two-mode approximation is employed to decompose the wave
function $\psi (x,t)$ over the minimum basis constituted by the ground state
$u_{0}$ and the first excited state $u_{1}$, associated to the eigenvalues $%
\omega _{0}$ and $\omega _{1}$, respectively. For this purpose, it is more
convenient to use a transformed orthonormal basis composed by wave functions
centered at the left and right wells, \textit{viz}., $\left\{ \varphi
_{L},\varphi _{R}\right\} $ $\equiv \left\{ (u_{0}-u_{1})/\sqrt{2}%
,(u_{0}+u_{1})/\sqrt{2}\right\} $. Without the loss of generality, $\varphi
_{L,R}$ are both chosen to be positive definite. Thus, the two-mode
approximation for the wave function is defined as
\begin{equation}
\psi (x,t)=c_{L}(t)\varphi _{L}(x)+c_{R}(t)\varphi _{R}(x),  \label{eq7}
\end{equation}%
where $c_{L}$ and $c_{R}$ are complex time-dependent amplitudes.
Substituting this into Eq. (\ref{eq1}), we obtain
\begin{equation}
\begin{split}
& i\dot{c}_{L}\varphi _{R}+i\dot{c_{R}}\varphi _{R}=(\Omega c_{L}-\mu
c_{L}-\omega c_{R})\varphi _{L}+(\Omega c_{R}-\mu c_{R}-\omega c_{L})\varphi
_{R} \\
& +s|c_{L}|^{2}(c_{L}\varphi _{L}+c_{R}\varphi _{R})\int \!K(x-x^{\prime
})\varphi _{L}^{2}(x^{\prime })dx^{\prime }+s|c_{R}|^{2}(c_{L}\varphi
_{L}+c_{R}\varphi _{R})\int \!K(x-x^{\prime })\varphi _{R}^{2}(x^{\prime
})dx^{\prime } \\
& +s[(c_{L}^{2}c_{R}^{\ast }+|c_{L}|^{2}c_{R})\varphi _{L}+(c_{L}^{\ast
}c_{R}^{2}+c_{L}|c_{R}|^{2})\varphi _{R}]\int \!K(x-x^{\prime })\varphi
_{L}(x^{\prime })\varphi _{R}(x^{\prime })dx^{\prime },
\end{split}
\label{eq8}
\end{equation}%
where the asterisk and overdot stand for the complex conjugate and time
derivative, while $\Omega \equiv (\omega _{0}+\omega _{1})/2$ and $\omega
\equiv (\omega _{1}-\omega _{0})/2$ are linear combinations of the two
lowest eigenvalues. Next, we project Eq. (\ref{eq8}) onto the single-well
states $\varphi _{L,R}$, which involves the following overlap integrals:
\begin{equation}
\begin{aligned} \eta_{0}&= \int\!\!\!\int
K(x-x')\varphi_{L}^{2}(x')\varphi_{L}^{2}(x)\:dx'dx,\\ \eta_{1}&=
\int\!\!\!\int K(x-x')\varphi_{L}^{2}(x')\varphi_{R}^{2}(x)\:dx'dx,\\
\eta_{2}&= \int\!\!\!\int
K(x-x')\varphi_{L}^{2}(x')\varphi_{L}(x)\varphi_{R}(x)\:dx'dx,\\ \eta_{3}&=
\int\!\!\!\int
K(x-x')\varphi_{L}(x')\varphi_{R}(x')\varphi_{L}(x)\varphi_{R}(x)\:dx'dx.
\end{aligned}  \label{eq9}
\end{equation}%
The equations hold if subscripts $L$ and $R$ are swapped, or the variable $x$
and $x^{\prime }$ are interchanged, due to the symmetry of kernel $K$. 
In Fig.~\ref{fig1} we show the values of the four integrals, $\eta _{0,1,2,3}$, 
as functions of parameter $\sigma $, 
for the three types of kernels defined in Eqs.~(\ref{eq4})-(\ref{eq6}). 
We note that $\eta _{2}$ and $\eta_{3}$ remain negligible for any value 
of $\sigma $, and $\eta _{0,1}$ are 
both positive when the kernel $K$ is positive definite. Naturally, when 
$\sigma $ is small, which corresponds to the limit of a nearly local
nonlinearity, $\eta _{0}$ is much larger that the other three overlap 
integrals, due to the weak overlapping of the single-well states $\varphi
_{L,R}$. \ As $\sigma $ increases, $\eta _{0}$ decreases while $\eta _{1}$
increases and, finally, they tend to become equal. Regarding the values of
these integrals, the upcoming analysis is conducted in two situations: $\eta
_{0}$ much larger than all others, and $\eta _{0}$ being on the same order
of magnitude as $\eta _{1}$. Some value of the $\sigma _{b}$ is to be fixed
as a boundary between the two cases. Since $\eta _{0}$ is always large, the
key point is to set up a rule for comparing $\eta _{1}$ with $\eta _{2,3}$.
To this end, we define $\eta _{\mathrm{rel}}=\eta _{1}-\max (|\eta
_{2}|,|\eta _{3}|)$ and the criterion is stated as follows: if $\eta _{%
\mathrm{rel}}\geq 0.01$, $\eta _{1}$ is taken into regard in the analysis of
the two-mode approximation; otherwise, $\eta _{1}$ is insignificant, and
only $\eta _{0}$ is kept. For the kernels considered in this context, the
criterion yields $\sigma _{b}=2.96$, $1.56$ and $1.91$ for the Gaussian,
exponential and Lorentzian kernels, respectively.

Thus, with $\sigma <\sigma _{b}$, all integrals $\eta _{1,2,3}$ are omitted,
and only $\eta _{0}$ is retained. Then, the projection of Eq. (\ref{eq8})
onto the two-mode set, $\varphi _{L,R}$, leads to the following ODE system:
\begin{equation}
\begin{aligned} i\dot{c}_{L} &= (\Omega-\mu)c_{L} - \omega c_{R}+ s\eta_{0}
|c_{L}|^{2}c_{L},\\ i\dot{c}_{R} &= (\Omega-\mu)c_{R} - \omega c_{L}+
s\eta_{0} |c_{R}|^{2}c_{R}. \end{aligned}  \label{eq10}
\end{equation}%
We then introduce the Madelung representation, $c_{L,R}=\rho
_{L,R}e^{i\theta _{L,R}}$ with real time-dependent $\rho _{L,R}$ and $\theta
_{L,R}$, and derive from Eq. (\ref{eq10}) a set of equations for $\rho
_{L,R} $ and $\theta _{L,R}$:
\begin{equation}
\left\{ \begin{aligned} \dot{\rho}_{L}&= \omega \rho_{R}\sin{\theta}\\
\dot{\theta}_{L}&= (\mu-\Omega)+\omega\frac{\rho_{R}}{\rho_{L}}\cos{\theta}
-s\eta_{0}\rho_{L}^{2}\\ \end{aligned}\right.  \label{eq11}
\end{equation}%
where $\theta \equiv \theta _{L}-\theta _{R}$ is the relative phase between
the two modes, and the equations for $\rho _{R}$ and $\theta _{R}$ are
obtained by interchanging subscripts $L$ and $R$,
and $\theta$ with $-\theta$, in Eq. (\ref{eq11})
directly. We focus on steady solutions to this system, i.e. $\dot{\rho}%
_{L,R}=\dot{\theta}_{L,R}=0$. Then, for solutions with nonzero amplitudes, $%
\theta $ may only take values $0$ or $\pi $, which correspond, respectively,
to equal or opposite signs of real stationary solutions for $c_{L}$ and $%
c_{R}$. Through a straightforward algebra, three stationary solutions are
thus found: the symmetric solution, with $\theta =0$ and $\rho
_{L,R}^{2}=(\mu -\omega _{0})/s\eta _{0}$, existing when $\mu >\omega _{0}$ (%
$\mu <\omega _{0}$) for $s=1$ ($s=-1$), i.e., for the repulsive (attractive)
long-range interactions; the antisymmetric solution, with $\theta =\pi $ and
$\rho _{L,R}^{2}=(\mu -\omega _{1})/s\eta _{0}$, existing when $\mu >\omega
_{1}$ ($\mu <\omega _{1}$) for $s=1$ ($s=-1$); and an \emph{asymmetric}
solution, with $\rho _{L,R}^{2}=(s(\mu -\Omega )\pm \sqrt{(\mu -\Omega
)^{2}-4\omega ^{2}})/2\eta _{0}$. As we assume $\eta _{0}>0$, for the
repulsive case ($s=1$) the asymmetric state exists only if $\theta =\pi $,
i.e., it bifurcates from the antisymmetric solution when $\mu >\Omega
+2\omega $. On the contrary, in the attractive case, it emerges from the
symmetric state ($\theta =0$) when $\mu <\Omega -2\omega $. These
conclusions agree with the general principles of the SSB theory, according
to which the attractive/repulsive nonlinearity breaks the
symmetry/anti-symmetry of solutions with equal numbers of particles in the
two wells \cite{Bergeman_2mode,todd,theo,shliz}.

Next, we consider the other situation, in which both $\eta _{0}$ and $\eta
_{1}$ are taken into consideration, while the other two are neglected (i.e.,
$\sigma \geq \sigma _{b}$). In this case, projecting Eq. (\ref{eq8}) onto $%
\varphi _{L,R}$ results in the following system:
\begin{equation}
\begin{aligned} i\dot{c}_{L} &= (\Omega-\mu)c_{L} - \omega c_{R}+
sc_{L}(|c_{L}|^{2}\eta_{0} + |c_{R}|^{2}\eta_{1}),\\ i\dot{c}_{R} &=
(\Omega-\mu)c_{R} - \omega c_{L}+ sc_{R}(|c_{R}|^{2}\eta_{0} +
|c_{L}|^{2}\eta_{1}). \end{aligned}  \label{eq12}
\end{equation}%
In terms of the Madelung representation, we transform Eqs. (\ref{eq12}) into
\begin{equation}
\left\{ \begin{aligned} \dot{\rho}_{L}&= \omega \rho_{R}\sin{\theta}\\
\dot{\theta}_{L}&= (\mu-\Omega)+\omega\frac{\rho_{R}}{\rho_{L}}\cos{\theta}
-s\eta_{0}\rho_{L}^{2}-s\eta_{1}\rho_{R}^{2}, \end{aligned}\right.
\label{eq13}
\end{equation}%
with the equations for $\rho _{R}$ and $\theta _{R}$ produced by swapping
subscripts $L$ and $R$, and $\theta$ with $-\theta$, as before. 
In this case, a set of three stationary
states are again obtained: the symmetric one, $\theta =0$ and $\rho
_{L,R}^{2}=(\mu -\omega _{0})/s(\eta _{0}+\eta _{1})$; the antisymmetric
state, $\theta =\pi $ and $\rho _{L,R}^{2}=(\mu -\omega _{1})/s(\eta
_{0}+\eta _{1})$. Finally, the asymmetric state, $\theta =\pi $ ($\theta =0$%
) for $s=1$ ($s=-1$) and $\rho _{L,R}^{2}=((\mu -\Omega )/s\eta _{0}\pm
\sqrt{(\mu -\Omega )^{2}/\eta _{0}^{2}-4\omega ^{2}/\Delta \eta ^{2}})/2$,
where $\Delta \eta =\eta _{0}-\eta _{1}>0$ for the kernels we consider,
exists at $\mu >\Omega +2\omega \eta _{0}/\Delta \eta $ ($\mu <\Omega
-2\omega \eta _{0}/\Delta \eta $) for $s=1$ ($s=-1$). The value of $\mu $ at
which the SSB\ bifurcation happens is tagged as $\mu ^{\mathrm{cr}}$. Since $%
\eta _{0}/\Delta \eta $ is an increasing function of $\sigma $, the value of
$\mu ^{\mathrm{cr}}$ increases (decreases) as $\sigma $ grows for $s=1$ ($%
s=-1$).

The two-mode approximation is a powerful means for identifying different
steady states of the underlying problem. The solutions found above are also
used as initial conditions in solving the full NLS equation (\ref{eq1}), as
reported in section \ref{secNum}.

\subsection{The bifurcation analysis}

The two-mode approximation strongly facilitates the qualitative analysis of
the SSB bifurcation, as well as exploring the system's dynamics. To proceed,
we define the population imbalance between the two wells,
\begin{equation}
z=(N_{L}-N_{R})/N=(|c_{L}|^{2}-|c_{R}|^{2})/N,  \label{eq14}
\end{equation}%
where $N_{L,R}=|c_{L,R}|^{2}\equiv \rho _{L,R}^{2}$, hence the total norm is
$N=N_{L}+N_{R}$. Together with the relative phase, $\theta =\theta
_{L}-\theta _{R}$, we eventually derive the following dynamical equations:
\begin{equation}
\left\{ \begin{aligned} \dot{z}&= 2\omega \sqrt{1-z^{2}}\sin{\theta}\\
\dot{\theta} &= -\frac{2\omega z \cos{\theta}}{\sqrt{1-z^{2}}} - s\eta N z.
\end{aligned}\right.  \label{eq15}
\end{equation}%
This form of the equations is relevant for both cases, when only $\eta _{0}$
or both $\eta _{0}$ and $\eta _{1}$ dominate, as discussed before. Note that
$\eta $ stands for $\eta _{0}$ in the former case, and for $\Delta \eta $ in
the latter one. Equations (\ref{eq15}) take the Hamiltonian form,
\begin{equation}
\left\{ \begin{aligned} \dot{z} &= &\:- \frac{\partial
\mathcal{H}}{\partial\theta},\\ \dot{\theta} &= &\: \frac{\partial
\mathcal{H}}{\partial z},\\ \end{aligned}\right.  \label{eq16}
\end{equation}%
with Hamiltonian
\begin{equation}
\mathcal{H}=2\omega \sqrt{1-z^{2}}\cos {\theta }-\frac{1}{2}s\eta Nz^{2}.
\label{eq17}
\end{equation}

Equations (\ref{eq15}) 
possess the stationary solutions, $%
(z_{1},\theta _{1})$ and $(z_{2},\theta _{2})$, with $z_{1}=z_{2}=0$, $%
\theta _{1}=0$, $\theta _{2}=\pi $, which represent the symmetric and the
antisymmetric solutions, respectively. Besides that, the asymmetric
stationary solutions may exist when $N\geq N^{\mathrm{cr}}$, with $N^{%
\mathrm{cr}}=|2\omega /\eta |$, taking the form of
\begin{equation}
\theta =\pi \;(\theta =0),\quad z^{2}=1-\frac{4\omega ^{2}}{\eta ^{2}N^{2}},
\label{eq18}
\end{equation}%
for $s=1$ ($s=-1$). The asymmetric solution emerges from the antisymmetric
(symmetric) one through a pitchfork bifurcation, in the case of the
defocusing (focusing) nonlinearity. Since $\eta =\eta _{0}$ for $\sigma
<\sigma _{b}$ and $\eta =\eta _{0}-\eta _{1}$ for $\sigma >\sigma _{b}$,
Fig. \ref{fig1} suggests that $\eta $ is, generally, a decreasing function
of $\sigma $, and consequently $N^{\mathrm{cr}}$ is increasing with respect
to $\sigma $. 
This way, the bifurcation takes place at larger value of $N$
when the the nonlocality range is wider, which is consistent with 
the conclusion 
concerning $\mu ^{\mathrm{cr}}$. 

\begin{figure}[tbph]
\centering
\includegraphics[width=.3\textwidth]{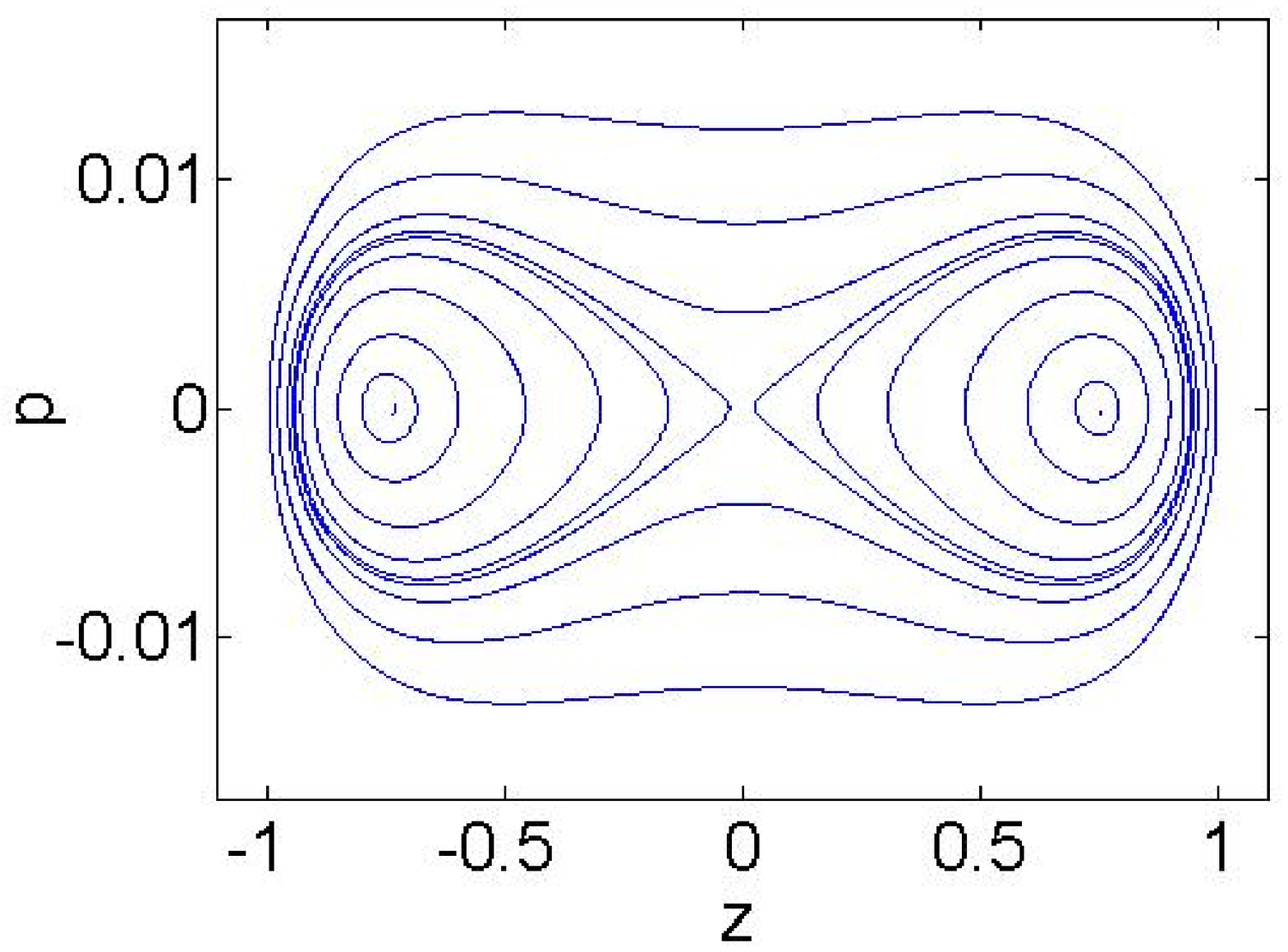} %
\includegraphics[width=.3\textwidth]{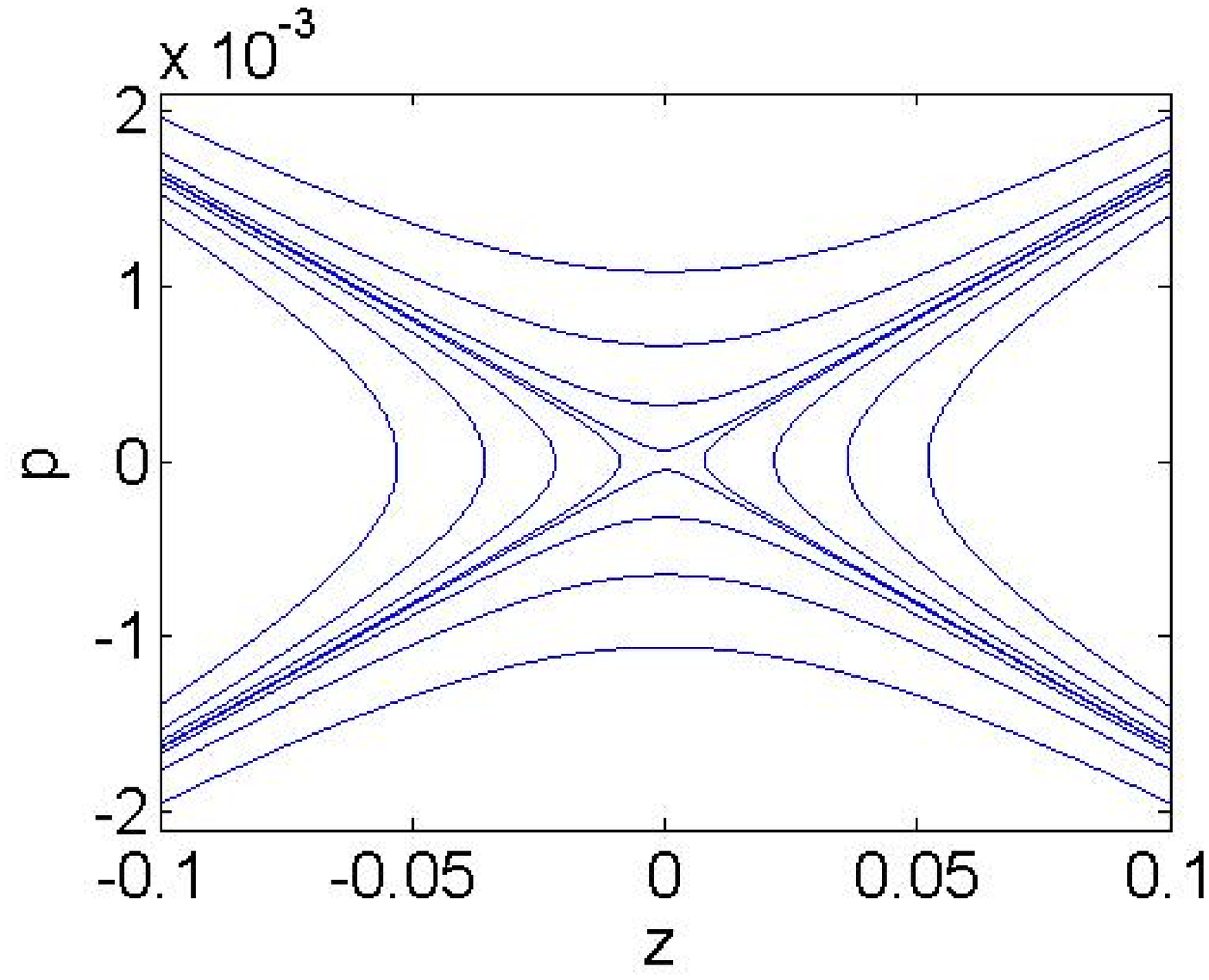} %
\includegraphics[width=.3\textwidth]{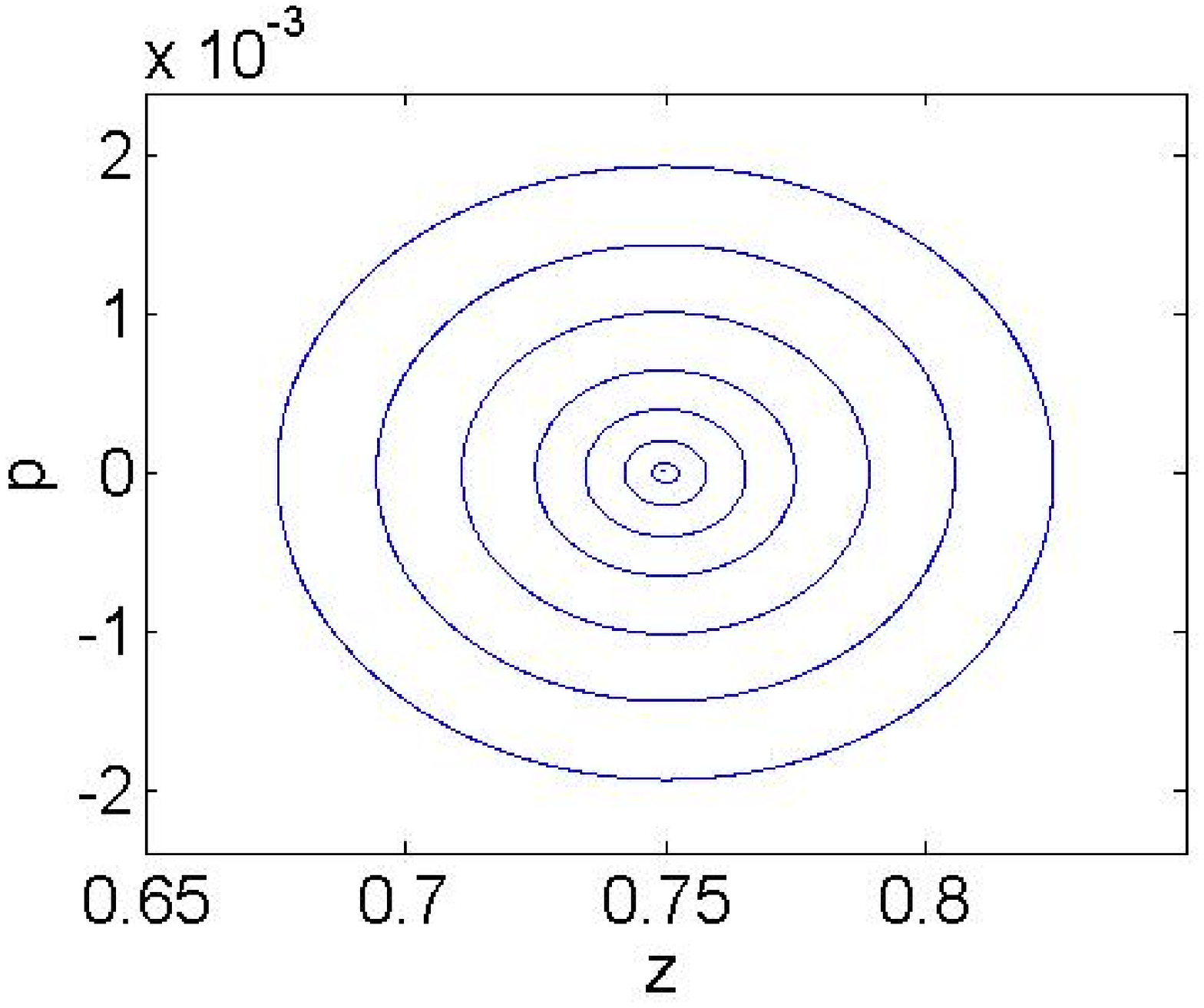}\newline
\caption{(Color online) Phase diagrams in the Gaussian-kernel model with
with $\protect\sigma =5.0$ and $N=0.5$, in which case $N^{\mathrm{cr}}=0.33$%
. The left panel is obtained from complete system (\protect\ref{eq20}),
displaying three fixed points, $(0,0)$ and $(\pm 0.7498,0  )$. The middle and
right panels are the phase planes of the linearized system near fixed points
$(0,0)$ (middle) and $(0.7498,0)$ (right).}
\label{fig2}
\end{figure}

Next, from Eq.~(\ref{eq15}) we derive the equation of motion,
\begin{equation}
\ddot{z}=-4\omega ^{2}z+|\eta |Nz\sqrt{4\omega ^{2}-4\omega ^{2}z^{2}-\dot{z}%
^{2}},  \label{eq19}
\end{equation}%
which leads to the system
\begin{equation}
\left\{ \begin{aligned} \dot{z} &= p,\\ \dot{p} &= -4\omega^{2}z +|\eta|N z
\sqrt{4\omega^{2}-4\omega^{2}z^{2}-p^{2}}. \end{aligned}\right.  \label{eq20}
\end{equation}%
When $N<N^{\mathrm{cr}}$, there is a unique stationary solution, $p=z=0$,
which is a fixed point of the center type. When $N>N^{\mathrm{cr}}$, the
origin, $(0,0)$, becomes a saddle, with another pair of fixed points
(centers), $p=0$, $z=\pm \sqrt{1-\dfrac{4\omega ^{2}}{|\eta |^{2}N^{2}}}$,
representing the asymmetric solutions. Figure \ref{fig2} shows the phase
space of system (\ref{eq20}), along with the linearization near the fixed
points for the example of the Gaussian kernel with $\sigma =5.0$ and $N=0.5$, 
in which case $N^{\mathrm{cr}}=0.33$.

\section{Numerical approach}

\label{secNum} 
\begin{figure}[tbph]
\centering
\includegraphics[width=.35\textwidth]{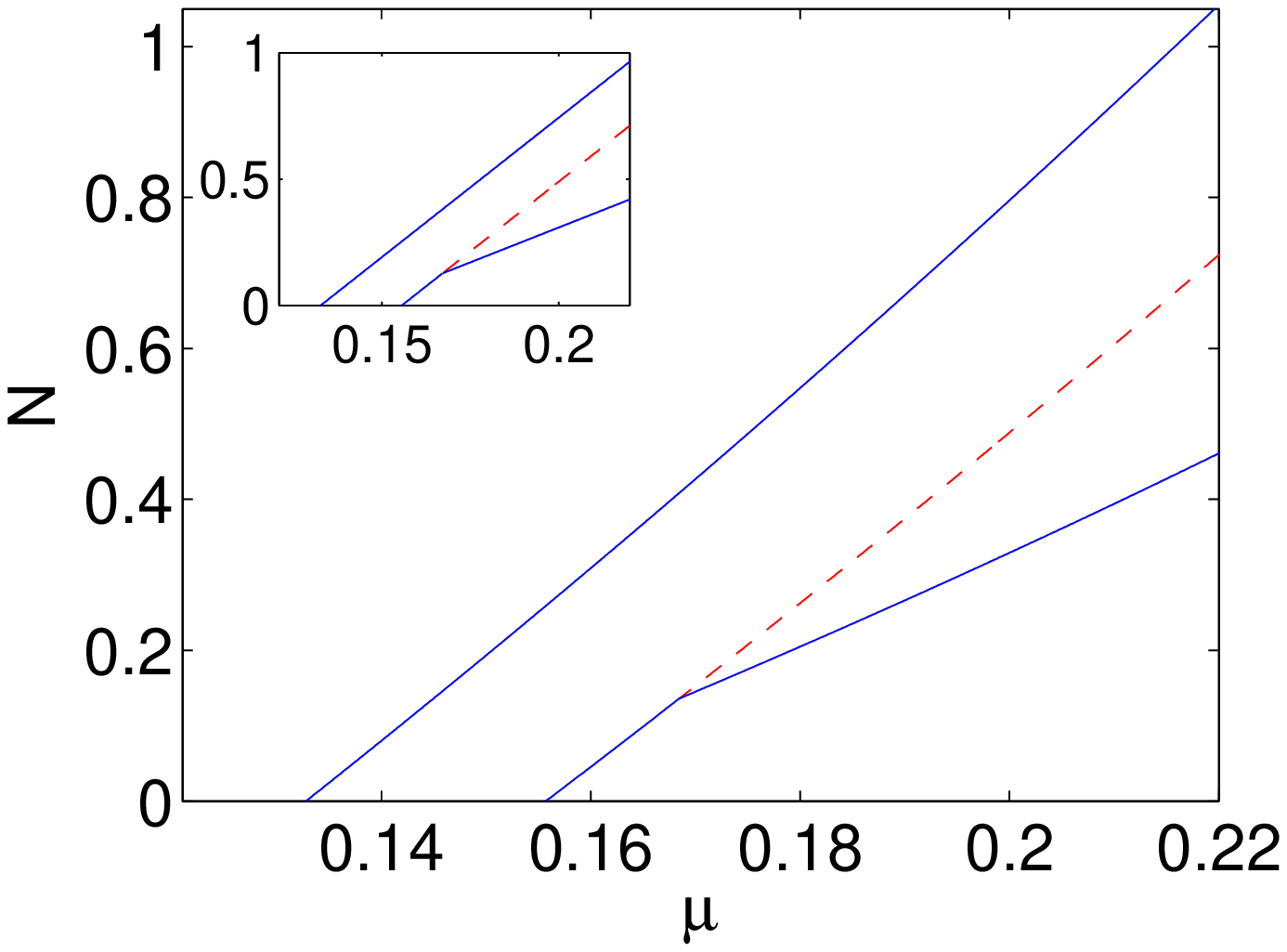} %
\includegraphics[width=.35\textwidth]{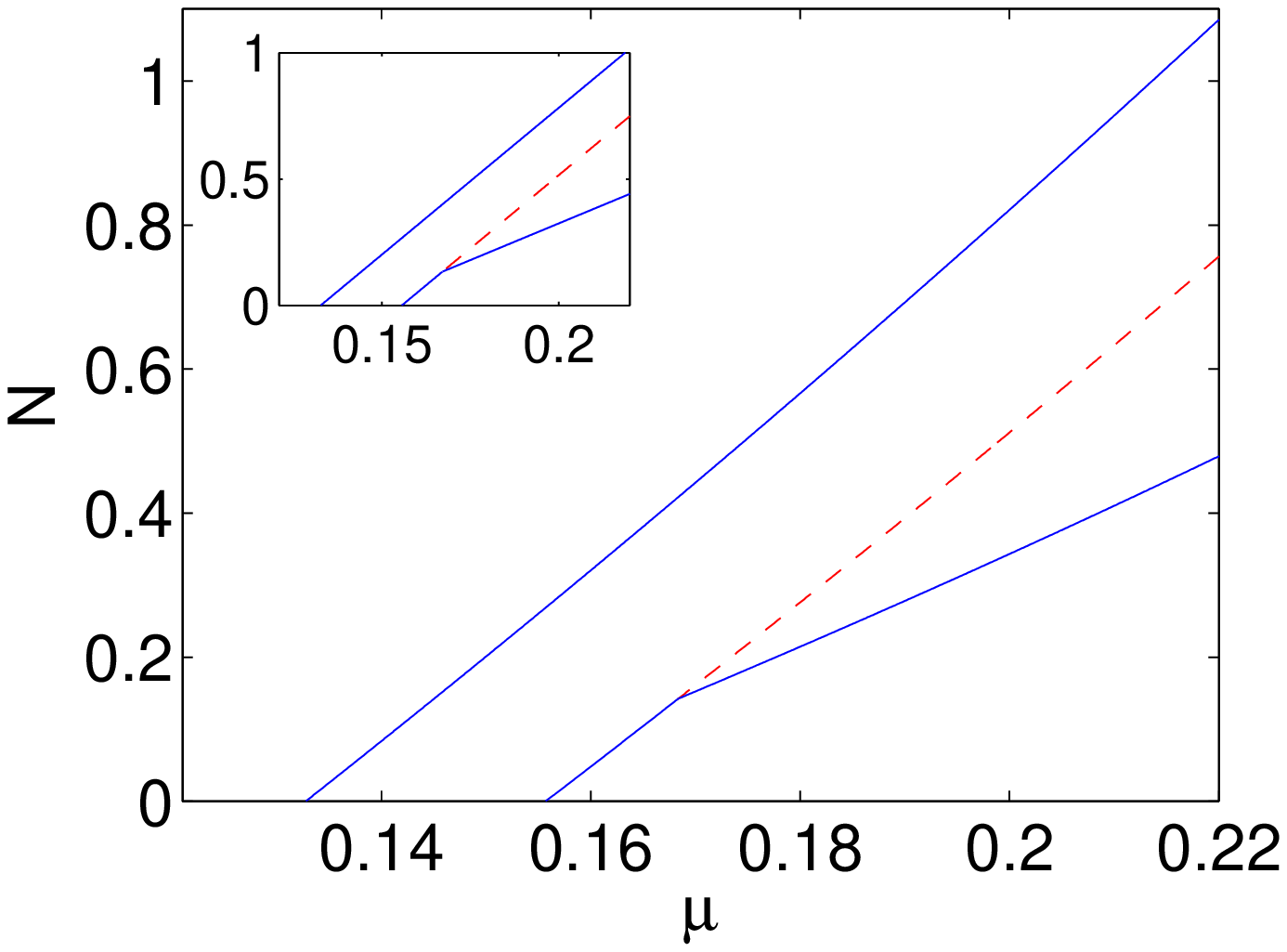}\newline
\includegraphics[width=.35\textwidth]{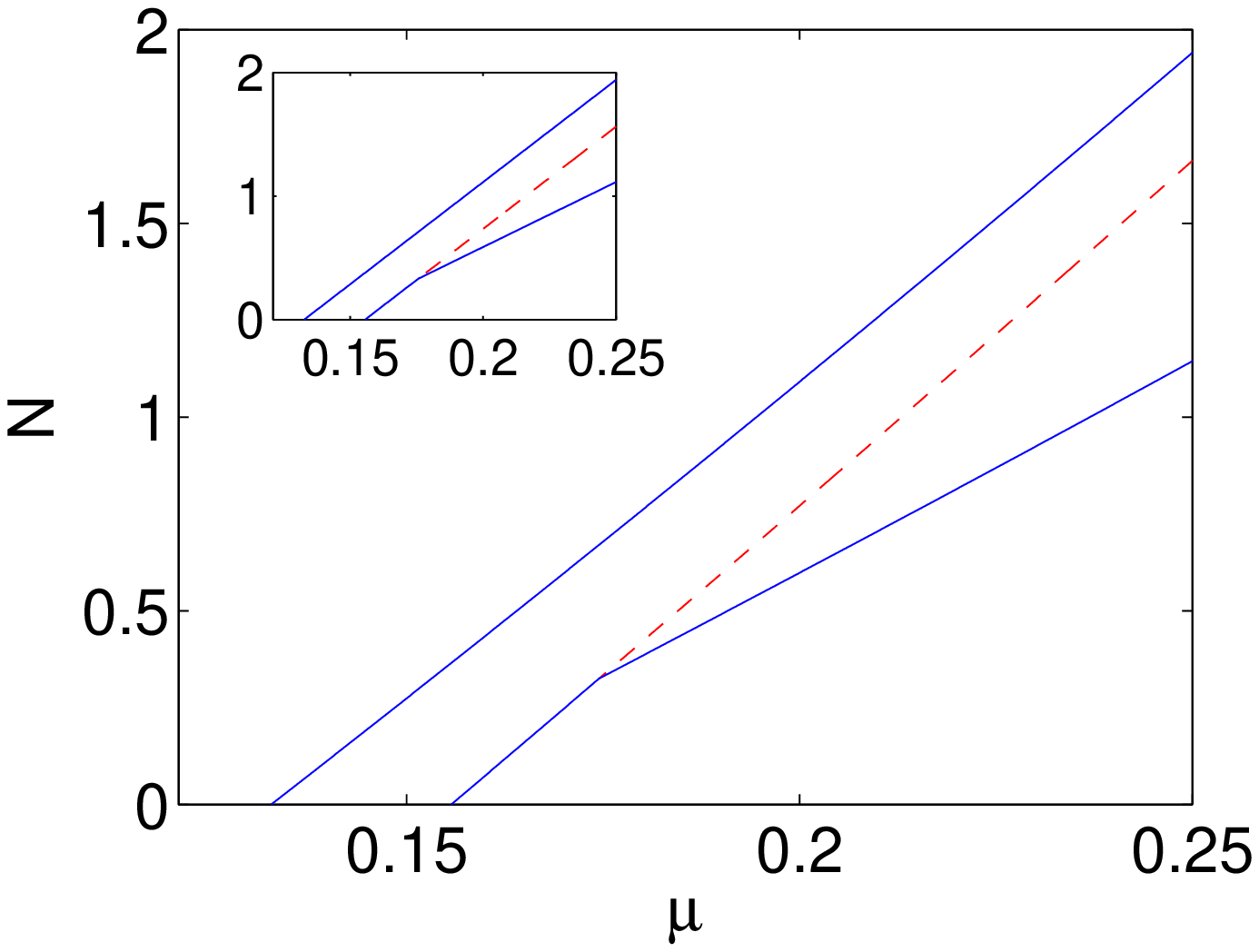} %
\includegraphics[width=.35\textwidth]{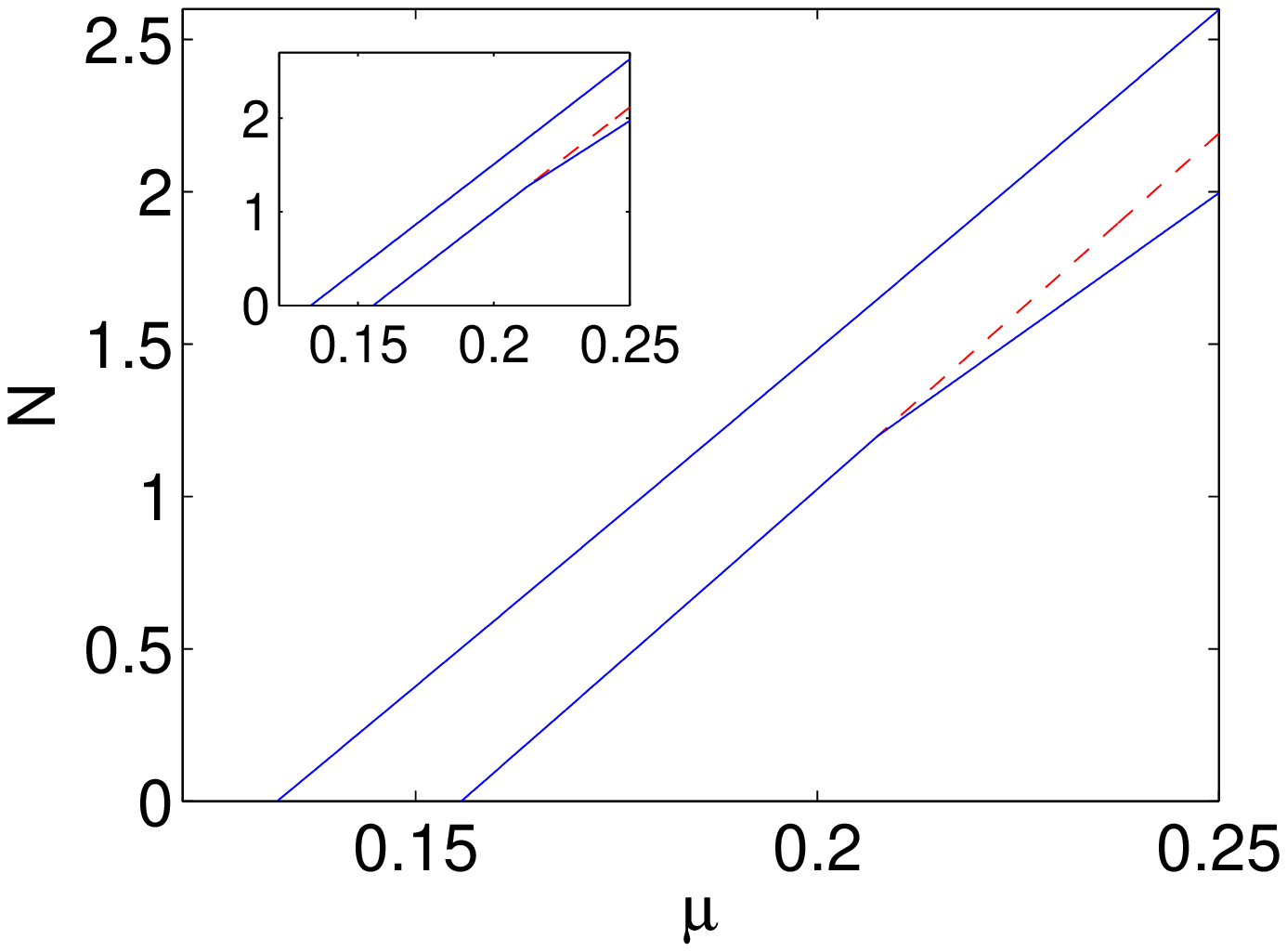}\newline
\caption{(Color online) The normalized norm, $N$, of the numerically found
stationary solutions of the underlying GP equation with the Gaussian kernel,
for the case of repulsive interaction ($s=1$), as a function of chemical
potential $\protect\mu $. The four panels correspond to cases with different
values of $\protect\sigma $, \textit{viz}., $\protect\sigma =0.1$ (top
left), $\protect\sigma =1$ (top right), $\protect\sigma =5$ (bottom left),
and, finally, $\protect\sigma =10$ (bottom right). The stationary solutions
predicted by the two-mode approximation for each case are shown in the small
plot in the top left corner of each panel. The blue solid lines and red
dashed lines denote stable and unstable solutions, respectively. }
\label{fig3}
\end{figure}

We first consider 
the repulsive
interaction case ($s=1$). Branches of stationary solutions to the full
partial differential equation (\ref{eq1}) are explored for all three kernels
defined in Eqs. (\ref{eq4})-(\ref{eq6}) with various values of parameter $%
\sigma$. The results are plotted in Figs. \ref{fig3}\,-\,\ref{fig6}, where
the solutions are expressed in terms of $N$, the number of atoms, as a
function of the chemical potential $\mu$. The stationary solutions are
sought by employing a fixed-point Newton-Raphson iteration onto a finite
difference scheme with $\Delta x=0.1$ and using the continuation of the
solutions with respect to $\mu$. 
The linear stability of each solution
$\psi_{0}$, is analyzed by
considering the standard linearization around it in the form $\psi(x,t)=%
\psi_{0}+\varepsilon[a(x)\mathrm{e}^{\lambda t}+b^{*}(x)\mathrm{e}^{\lambda^{*}t}]$. 
The relevant linear eigenvalue problem
is written as
\begin{equation}
\begin{pmatrix}
L_{1} & L_{2} \\
-L^{*}_{2} & -L^{*}_{1}%
\end{pmatrix}%
= i \lambda
\begin{pmatrix}
a \\
b%
\end{pmatrix}%
,  \label{eqLEP}
\end{equation}%
where operators $L_{1}$, $L_{2}$ are defined as
\begin{equation}
\begin{aligned} L_{1}\phi &= \left[-\frac{1}{2} \partial^{2}_{x}+V-\mu +
s\int_{-\infty }^{\infty}K\,(x-x^{\prime })|\,\psi_{0}(x^{\prime })|^{2}
\,dx^{\prime }\,\right]\,\phi + s\int_{-\infty }^{\infty}K\,(x-x^{\prime })
\psi_{0}(x^{\prime })\psi^{*}_{0}(x^{\prime}) \phi(x^{\prime})\,dx^{\prime
},\\ L_{2}\phi &= s\int_{-\infty }^{\infty}K\,(x-x^{\prime })
\psi_{0}(x^{\prime })\psi^{*}_{0}(x^{\prime}) \phi(x^{\prime})\,dx^{\prime }
\end{aligned}  \label{eqL12}
\end{equation}
for any function $\phi$. The stationary state is called unstable if there
exist any eigenvalues $\lambda$ with $\Re(\lambda)\neq0$, otherwise it is
stable (i.e. all corresponding eigenvalues are purely imaginary).

We present the results of Gaussian kernel in detail as an example. In each
panel (Fig. \ref{fig3}), the solid blue line with highest value of $N$ for
any $\mu $ among the three branches is the symmetric stationary solution;
the continuation of it to the linear limit ($N\rightarrow 0$) shows that it
starts from $\mu =\omega _{0}$ (the eigenvalue associated to the ground mode
of the underlying linear system), and it is stable for any $\mu $. The branch
with a part of it 
denoted by dashed red line is the antisymmetric solution,
arising from the linear limit at $\mu =\omega _{1}$, corresponding to the
first excited state. It starts as a stable state from the linear mode
(denoted by solid blue line) and, as $\mu $ increases, 
after some critical
point $\mu ^{\mathrm{cr}}$, it is destabilized due to the emergence of the asymmetric
branch through a supercritical pitchfork; 
notice that the asymmetric branch remains stable after it occurs. 
This bifurcation 
was predicted by the two-mode analysis in the previous section.

\begin{figure}[tbph]
\centering
\includegraphics[width=.35\textwidth]{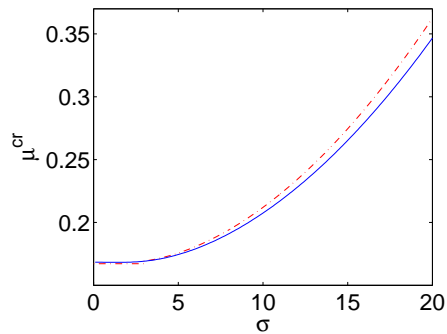}
\caption{(Color online) The critical value of the chemical potential, $%
\protect\mu ^{\mathrm{cr}}$, at which the supercritical pitchfork
bifurcation takes place in the case of the self-repulsive nonlinearity with
the Gaussian kernel, as a function of $\protect\sigma $. The blue solid line
and the red dashed-dotted line denote the numerically obtained values and
their counterparts predicted by the two-mode approximation, respectively.
Note that there is a small jump at $\protect\sigma =2.96$ on the
dashed-dotted line, due to the definition if the border between the two
situations [described by Eqs. (\protect\ref{eq10}) and (\protect\ref{eq12}),
respectively] in the two-mode analysis.}
\label{fig4}
\end{figure}

The four panels of Fig. \ref{fig3} are the bifurcation diagrams of the model
with the Gaussian kernel, for $\sigma =0.1,1,5$ and $10$. The main parts of
the panels display the numerically found solutions of the full system, while
the small plot in each panel shows the corresponding analytical solutions
predicted by the two-mode approximation, as obtained in Section \ref%
{secAnaly}. The numerical and analytical branches of the solutions
demonstrate good agreement in all the four cases. Recall that we\ have
obtained the critical values at which the bifurcation takes place, $\mu ^{%
\mathrm{cr}}$ (or equivalently $N^{\mathrm{cr}}$), in 
an analytical form.
As for the Gaussian kernel (with the chosen border at $\sigma _{b}=2.96$), $%
\sigma =0.1$ and $\sigma =1$ are categorized as belonging to the first case,
with solely $\eta _{0}$ taken into account, among all the overlap integrals.
The analytically predicted value is $\mu ^{\mathrm{cr}}=0.1671$ (since $\mu
^{\mathrm{cr}}=\Omega +2\omega $), while its numerically found counterpart
is $\mu ^{\mathrm{cr}}=0.1684$ for both $\sigma =0.1$ and $1$. On the other
hand, $\sigma =5$ and $\sigma =10$ pertain to the the second case, in which
both $\eta _{0}$ and $\eta _{1}$ are kept. 
In this case, the analytical
prediction is $\mu ^{\mathrm{cr}}=\Omega +2\omega \eta _{0}/\Delta \eta $,
which depends on $\eta _{0,1}$ and thus varies for different values of $%
\sigma $. The predicted values of $\mu ^{\mathrm{cr}}$ is $0.1757$ and $%
0.2120$, for $\sigma =5$ and $10$, respectively, while the respective
numerical values are $\mu ^{\mathrm{cr}}=0.1745$ and $0.2075$. Despite a
small discrepancy between the two groups of the values (numerical versus
analytical), the analytical results still succeed in predicting the trend of
the behavior of $\mu ^{\mathrm{cr}}$, \textit{viz}., $\mu ^{\mathrm{cr}}$
increases with the growth of $\sigma $, or, in other words, the SSB\
bifurcation takes place at higher values of $N^{\mathrm{cr}}$, as shown in
Fig. \ref{fig4}.

\begin{figure}[tbph]
\centering
\includegraphics[width=.35\textwidth]{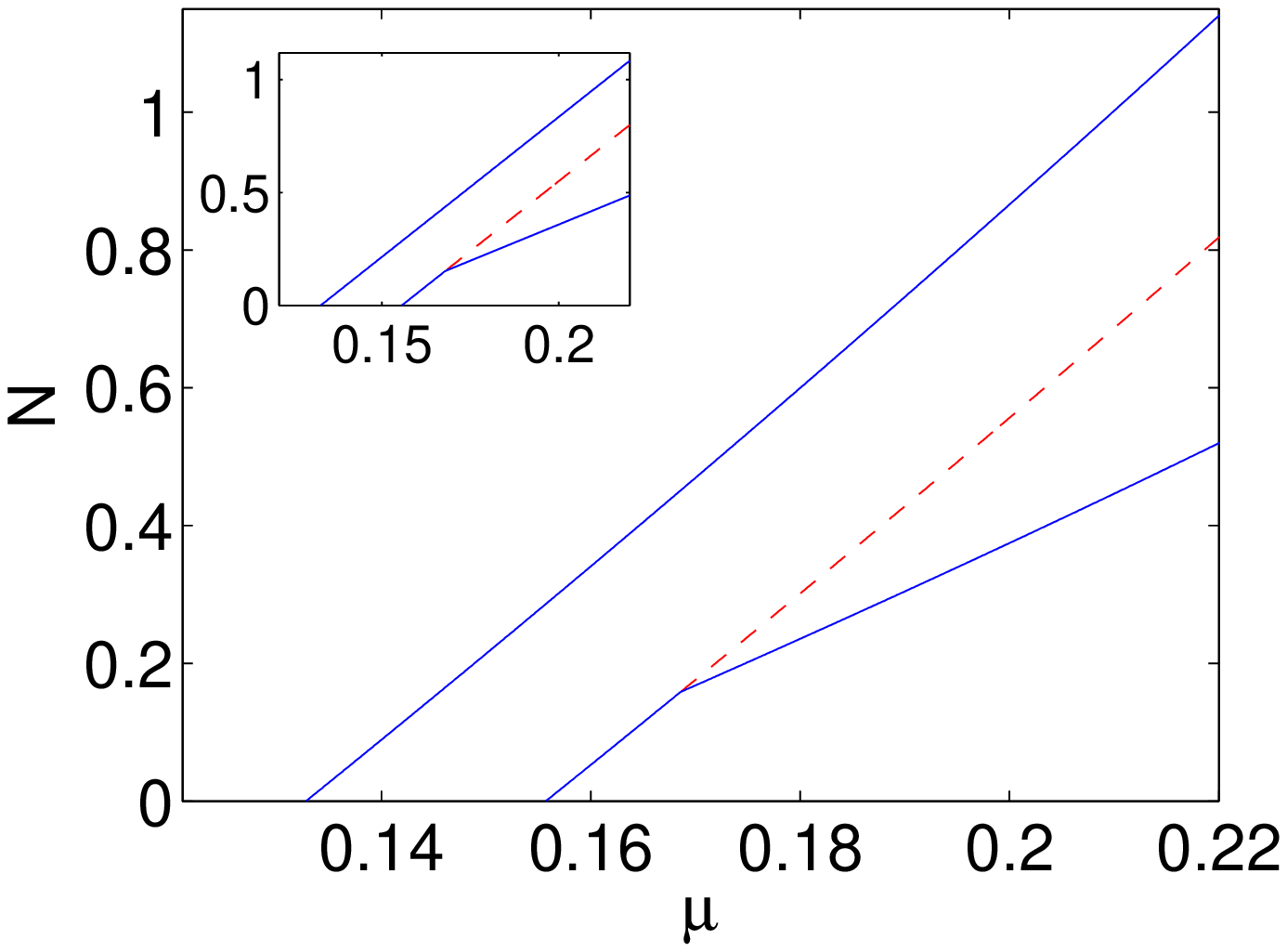} %
\includegraphics[width=.35\textwidth]{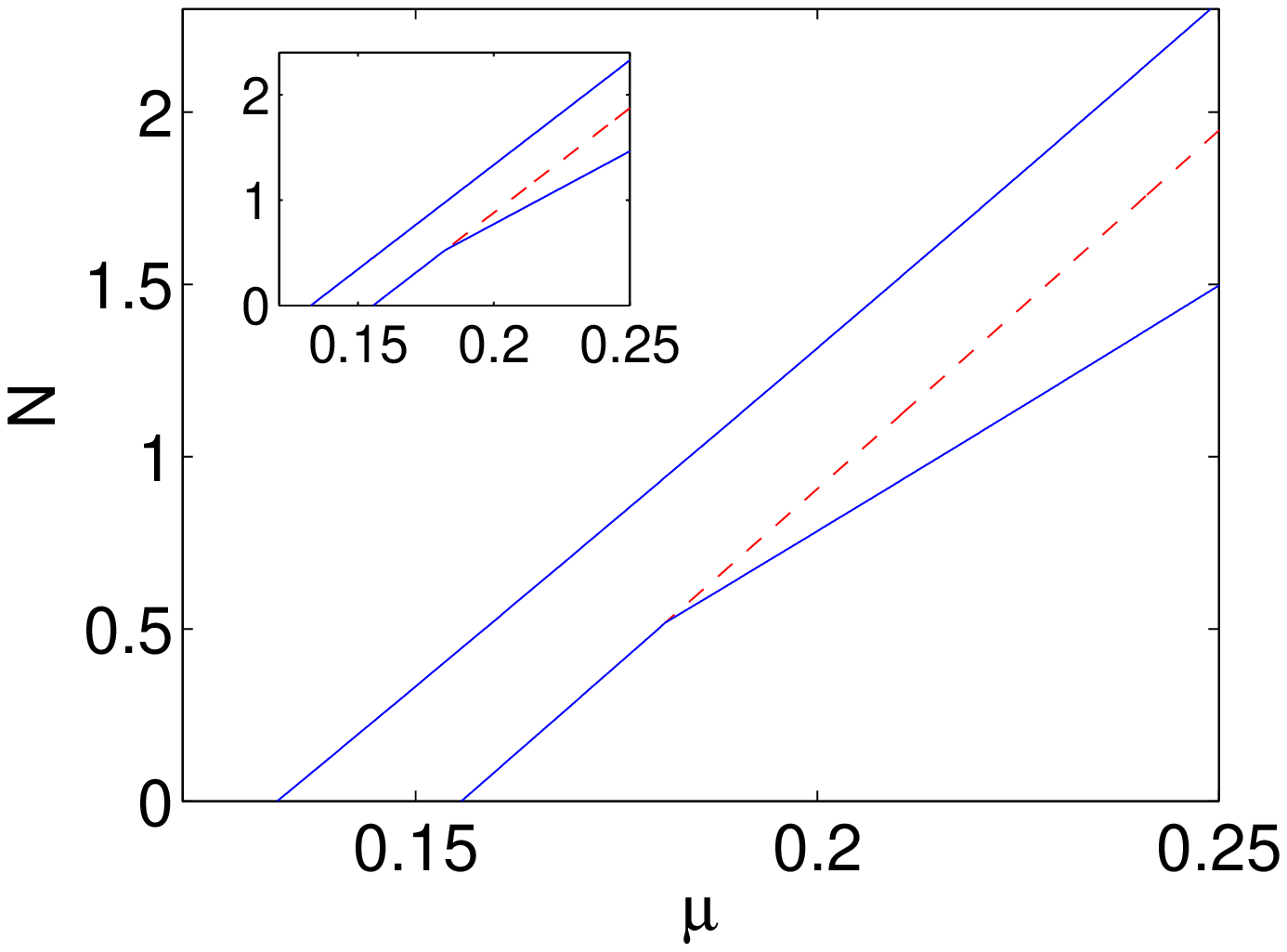}\newline
\caption{(Figure online) Norm $N$ of the numerically found solutions and
their counterparts predicted by the two-mode approximation (shown in the
corners) for the case of the self-repulsive nonlinearity with the
exponential kernel, as a function of $\protect\mu $. Here and in the next
figure, the value of $\protect\sigma$ is chosen as $\protect\sigma=1$
(left), $\protect\sigma=5$ (right), and the notation is the same as in Fig.
\protect\ref{fig3}.}
\label{fig5}
\end{figure}

\begin{figure}[tbph]
\centering
\includegraphics[width=.35\textwidth]{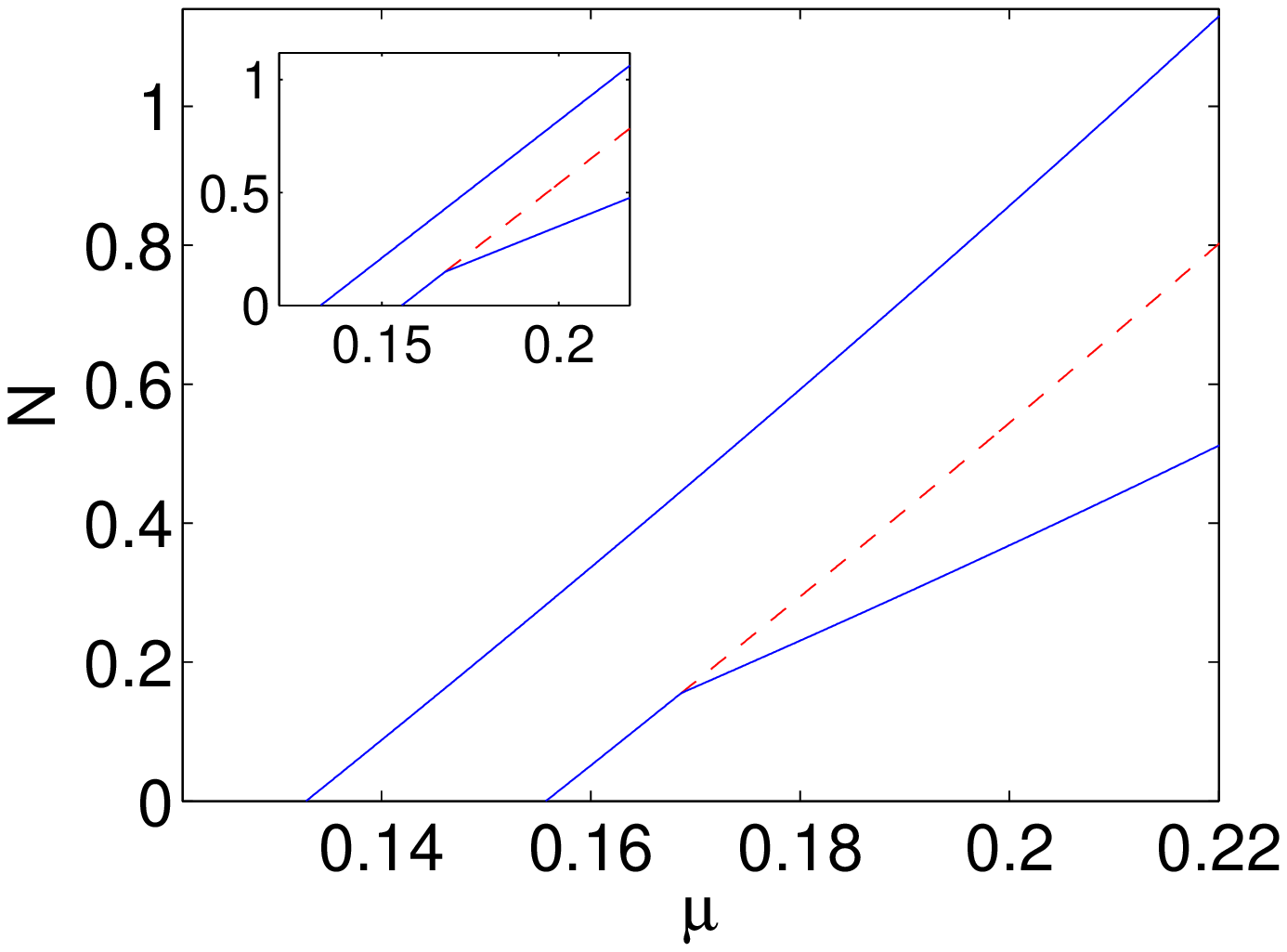} %
\includegraphics[width=.35\textwidth]{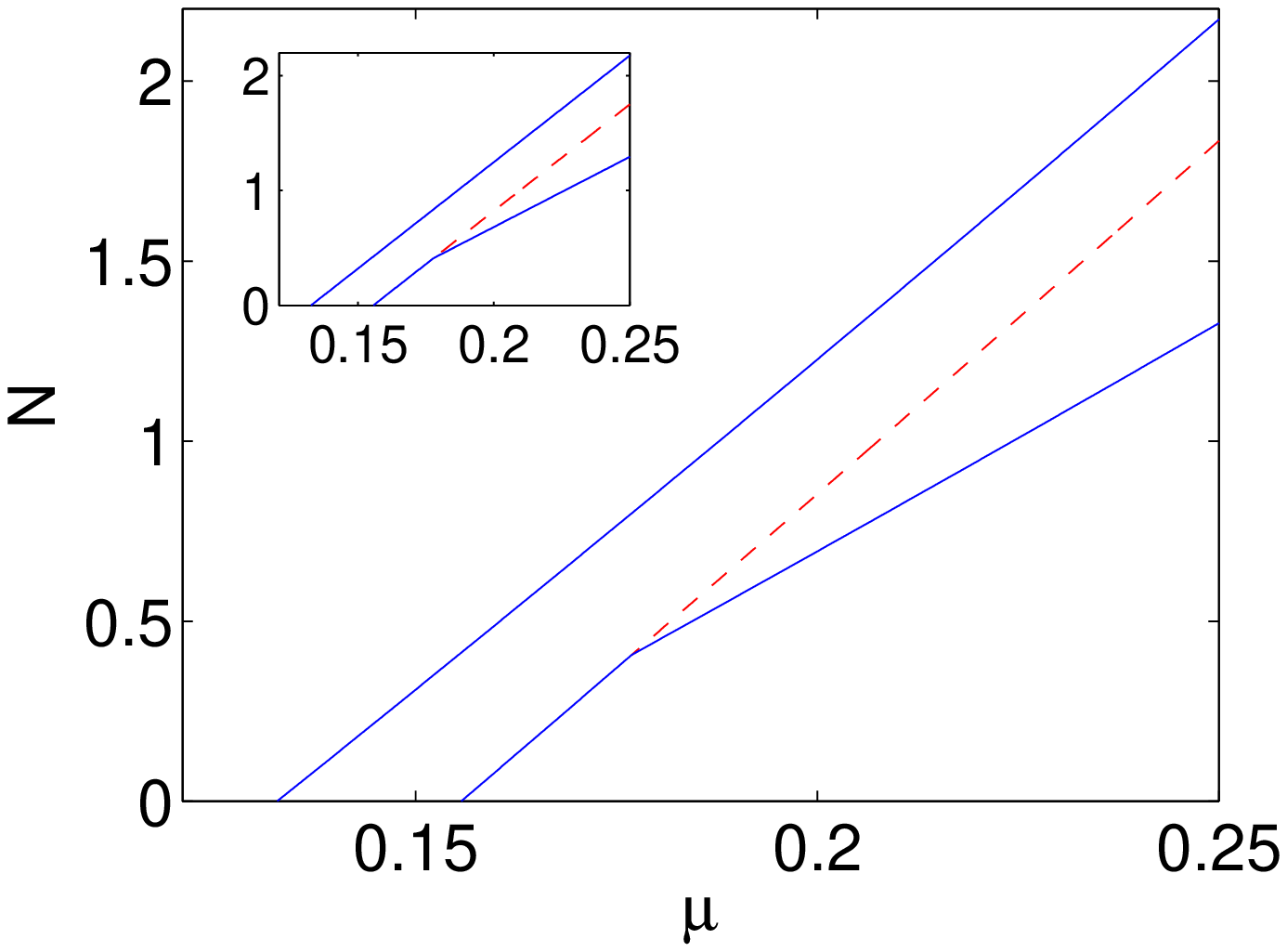}\newline
\caption{(Color online) The norm of the numerical and approximate analytical
solutions for the case of the self-repulsive nonlinearity with the
Lorentzian kernel, as a function of $\protect\mu $.}
\label{fig6}
\end{figure}

Next we consider cases with the long-range interaction based on the other
two kernels, \textit{viz}., the exponential and Lorentzian ones, given by
Eq.~(\ref{eq5}) and (\ref{eq6}), respectively. The corresponding bifurcation diagrams,
similar to the case of the Gaussian kernel, are presented in Fig.~\ref{fig5}
and Fig. \ref{fig6}, with the same notations as adopted above.

\begin{figure}[tbph]
\centering
\includegraphics[width=.35\textwidth]{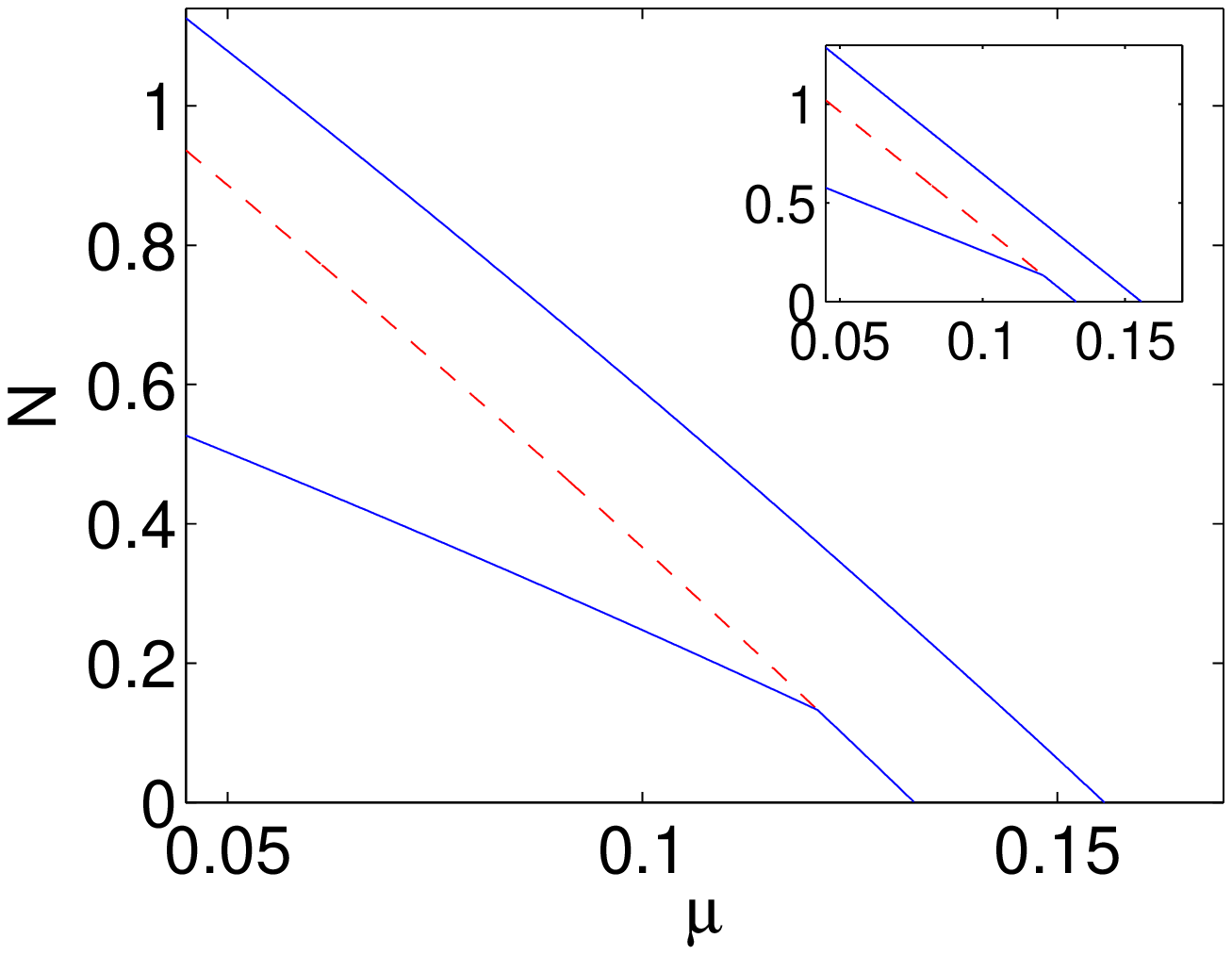} %
\includegraphics[width=.35\textwidth]{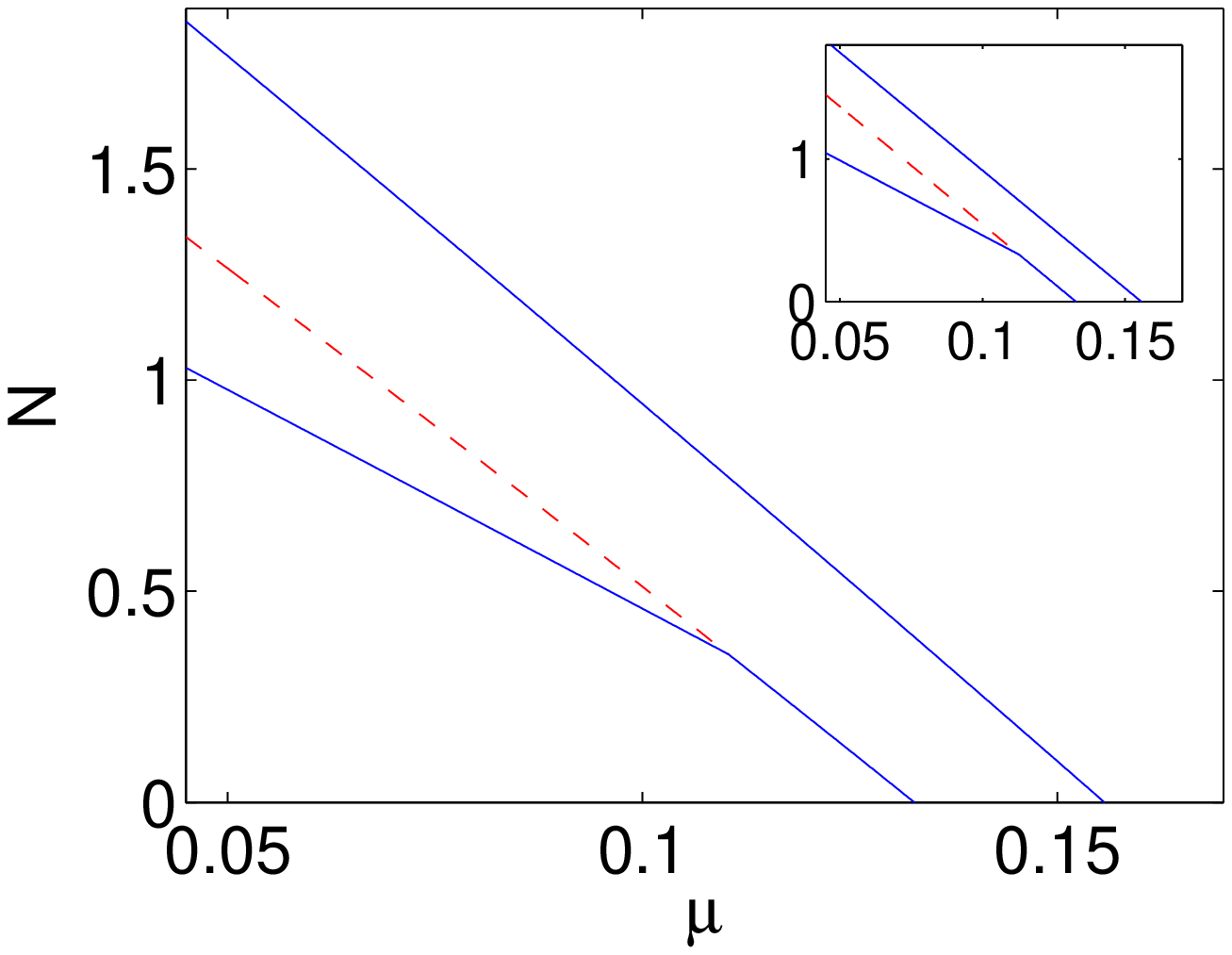}\newline
\caption{(Color online) Norm $N$ of the numerical solutions, together with
their analytical counterparts (shown in the small plots on the top right
corners), for the case of the self-attractive nonlinearity ($s=-1$), as a
function of $\protect\mu $. The Gaussian kernel is taken, with $\protect%
\sigma =1$ (left) and $\protect\sigma =5$ (right). The notation is the same
as in Fig. \protect\ref{fig3}. }
\label{fig7}
\end{figure}

Lastly, we also briefly discuss the self-attractive case, with $s=-1$ in 
Eq.~(\ref{eq1}). Two examples are 
selected, using the Gaussian kernel with $\sigma
=1$ and $5$, to help illustrating the similarities and differences with the
self-repulsive case. The complete bifurcation diagrams are plotted, for this
case, in Fig. \ref{fig7}, with the notation similar to that introduced above.

In this case, the two branches arise from the linear modes at $\mu =\omega
_{0}$ and $\omega _{1}$, pertaining to the symmetric and antisymmetric
stationary solutions, which exist for $\mu <\omega _{0}$ or $\mu
<\omega _{1}$, respectively. This time, the supercritical SSB\ pitchfork
bifurcation occurs on the symmetric branch, leading to the emergence of the
asymmetric state. The pitchfork is found at $\mu ^{\mathrm{cr}}=0.1213$ and $%
0.1128$ for $\sigma =1$ and $5$ respectively, in good agreement with the
critical values predicted by the two-mode approximation, which are $0.1212$
and $0.1104$. In each panel of Fig. \ref{fig7}, the diagram produced by the
two-mode approximation, corresponding to its numerical counterpart, is
displayed in the top right corner, exhibiting a very good agreement between
the two.

\section{Competing short- and long-range interactions}

In this section, we consider a model illustrating effects of the competition
of the long-range interactions with local ones, based on the following GP
equation, 
\begin{equation}
i\partial _{t}\psi +\mu \psi =L\psi +s\left[ \int_{-\infty }^{\infty
}K\,(x-x^{\prime })|\,\psi (x^{\prime })|^{2}\,dx^{\prime }\,\right] \,\psi
+g|\psi |^{2}\psi ,  \label{eqCOMP}
\end{equation}%
where coefficient $g$ accounts for the contact nonlinearity. 
The interactions are fully attractive or repulsive if $s
$ and $g$ are both negative or positive, respectively. However, in such
cases the results turn out to be similar to those reported above for the
nonlocal nonlinearity. A more interesting case arises for $sg<0$, i.e., for
the \textit{competing} interactions. Effects of the competition on 1D
solitons were recently studied in detail in Ref.~\cite{us}.

\begin{figure}[tbph]
\centering
\includegraphics[width=.35\textwidth]{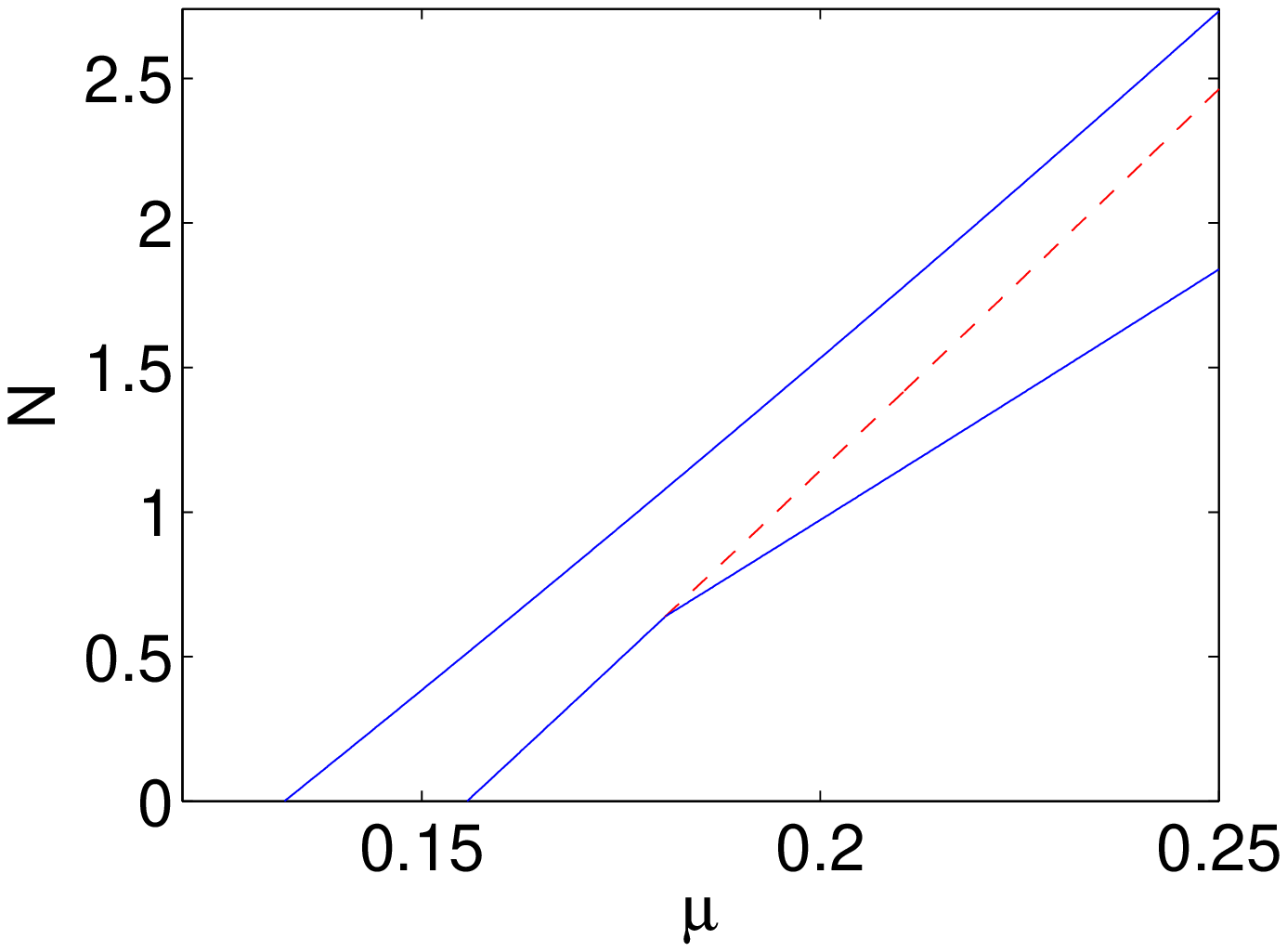} %
\includegraphics[width=.35\textwidth]{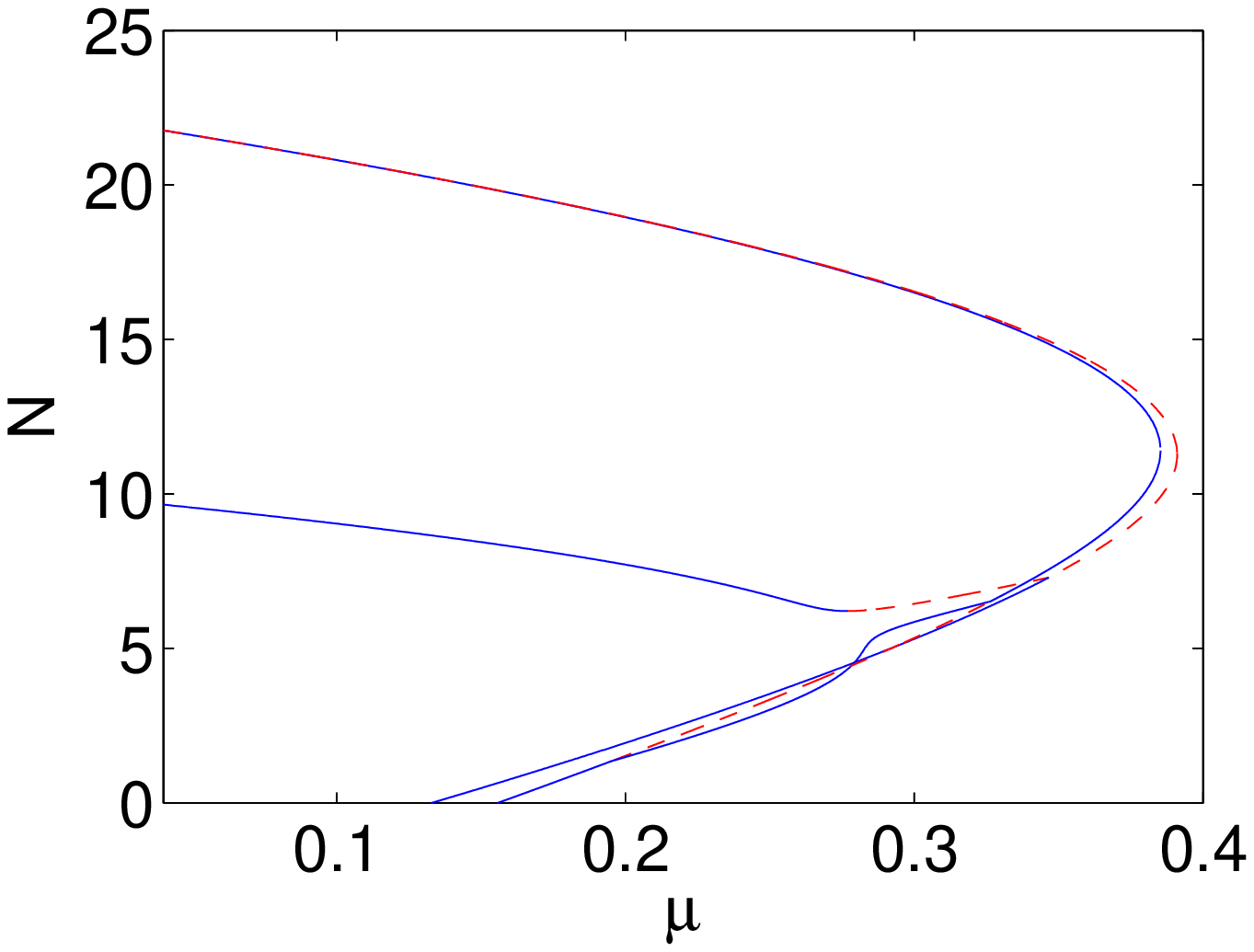}\newline
\includegraphics[width=.35\textwidth]{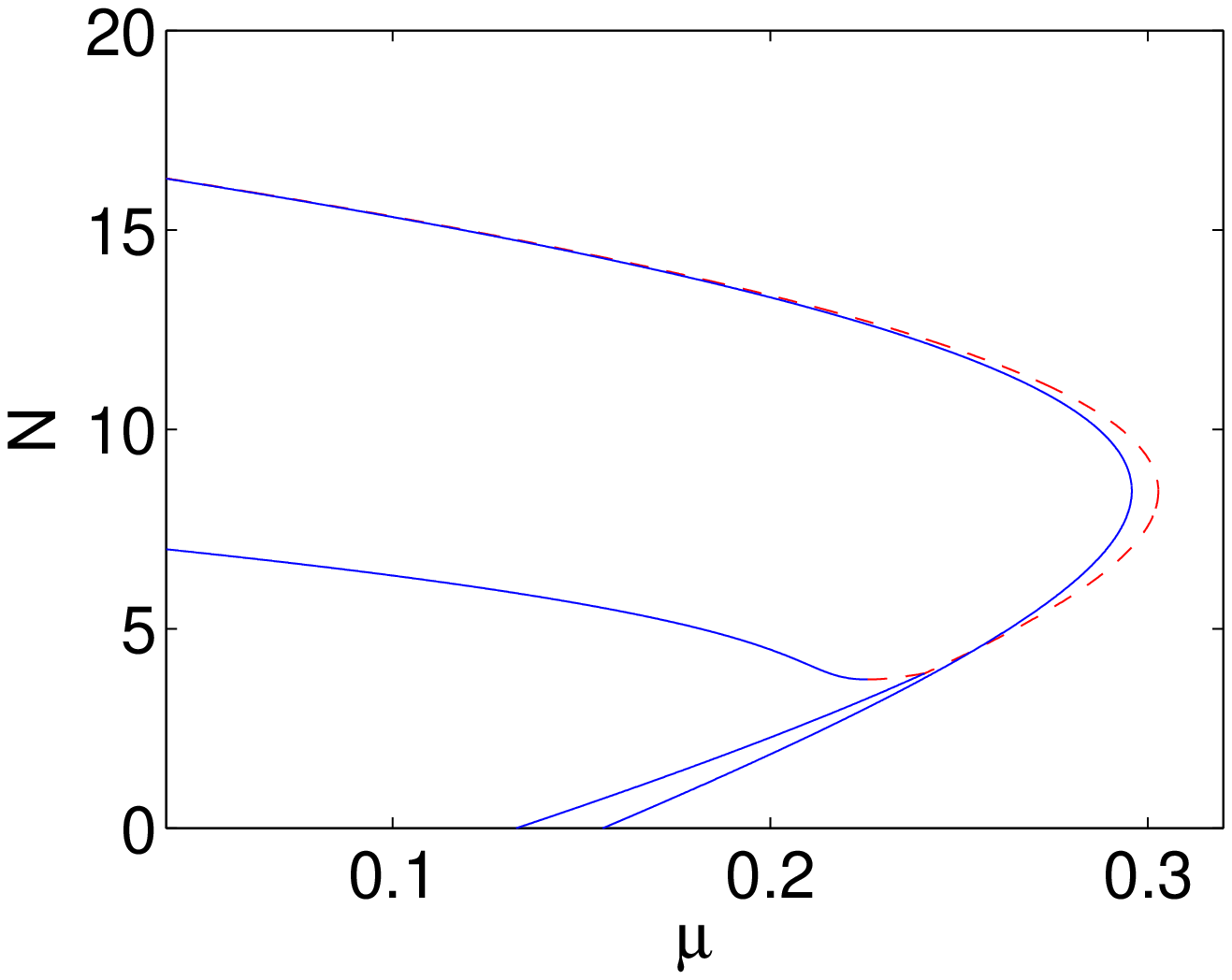} %
\includegraphics[width=.35\textwidth]{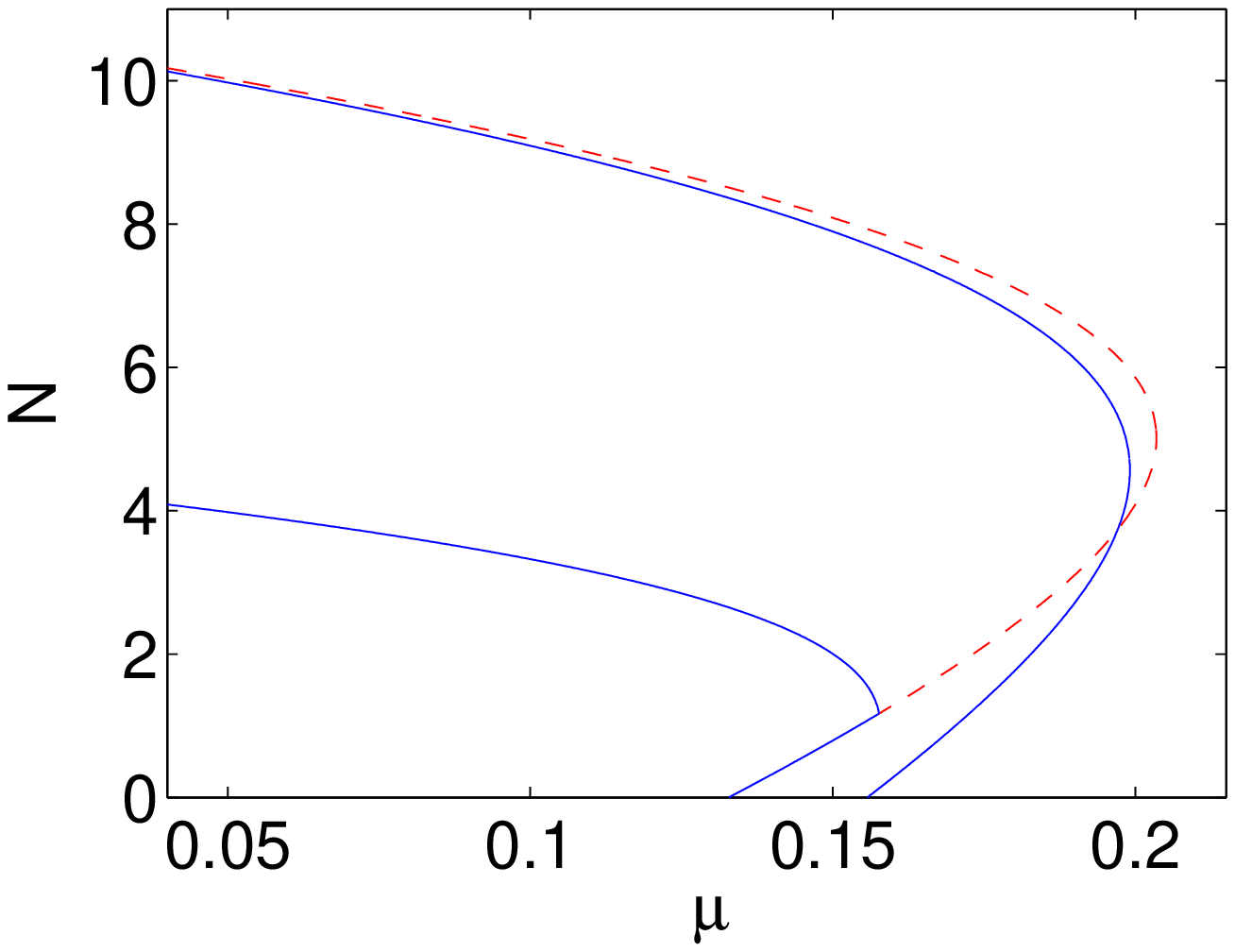}\newline
\caption{(Color online) Norm $N$ of the numerically found solutions in the
case of the competing interactions for Gaussian kernel (\protect\ref{eq4})
with $\protect\sigma =5$, as a function of $\protect\mu $. The coefficient
of the long-range interactions is $s=1$, while the local-interaction
coefficient is $g=-0.2$ (top left), $-0.3$ (top right), $-0.35$ (bottom
left) and $-0.45$ (bottom right). The notation is the same as in Fig.
\protect\ref{fig3}. }
\label{fig8}
\end{figure}

We use the analysis presented in the previous section as a starting point
for our considerations here. Thus, we fix the long-range interaction
coefficient $s=\pm 1$, and then vary the local interaction coefficient, $g$
from $0$ to $-1$ ($g<0$ implies the local attraction), using Gaussian kernel
(\ref{eq4}) with $\sigma =5$ and $s=1$ as a case example. As the attractive
local interactions grow stronger, interesting phenomena emerge in the
development of the corresponding bifurcation diagram, as is shown in Fig.~%
\ref{fig8}.

When $g$ is close to zero, the three branches of stationary states are
similar to those in the case of 
pure long-range interactions, as shown in
the bottom left panel of Fig. \ref{fig3}. An example for $g=-0.2$ is
presented in the top left panel of Fig. \ref{fig8}, where the norm $N$ of each
branch is larger at any value of $\mu $, in comparison to the case of $g=0$.
Also, the bifurcation takes place at a larger critical value, $\mu ^{\mathrm{%
cr}}=0.1805$.

\begin{figure}[tbph]
\centering
\includegraphics[width=.3\textwidth]{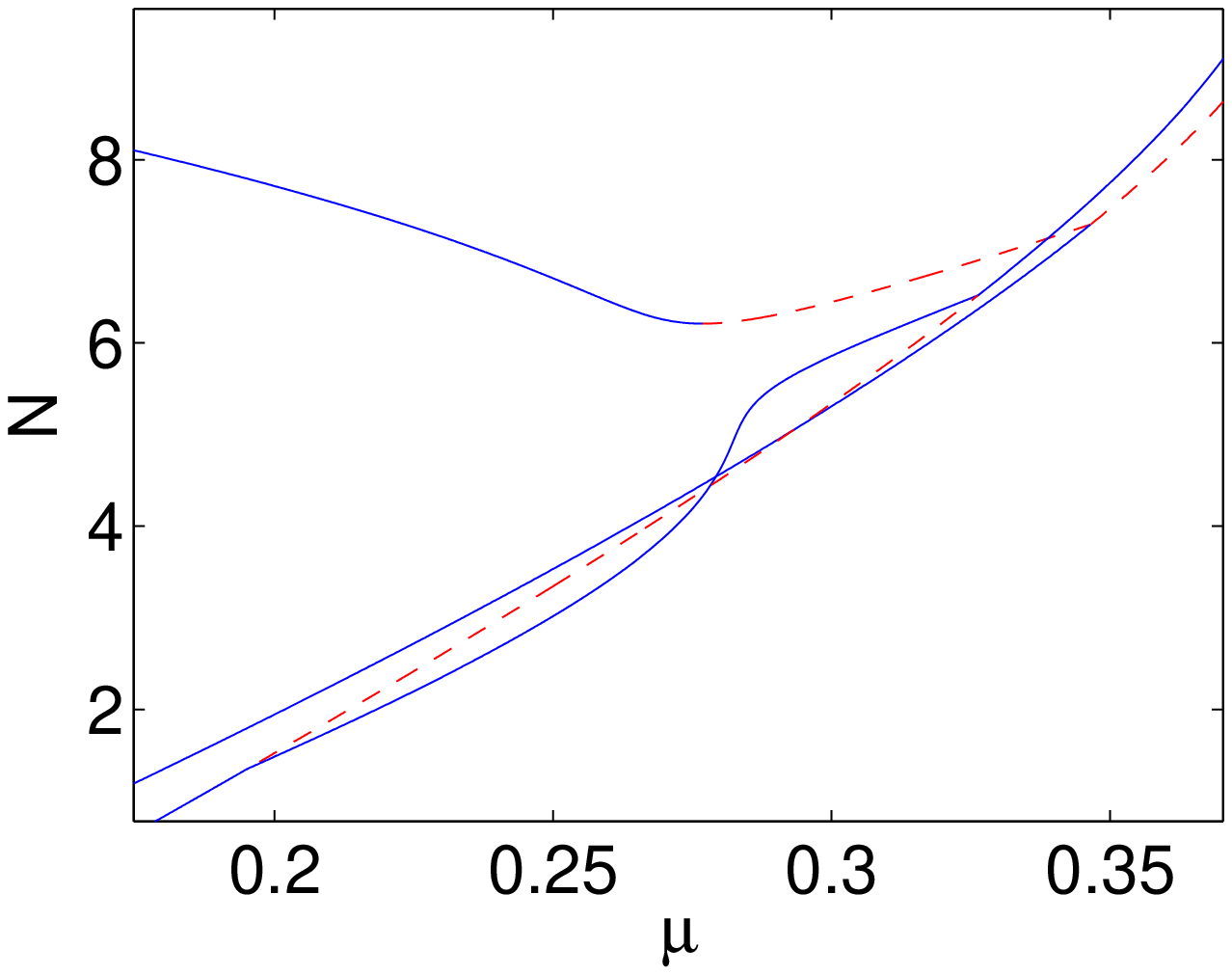}%
\includegraphics[width=.3\textwidth]{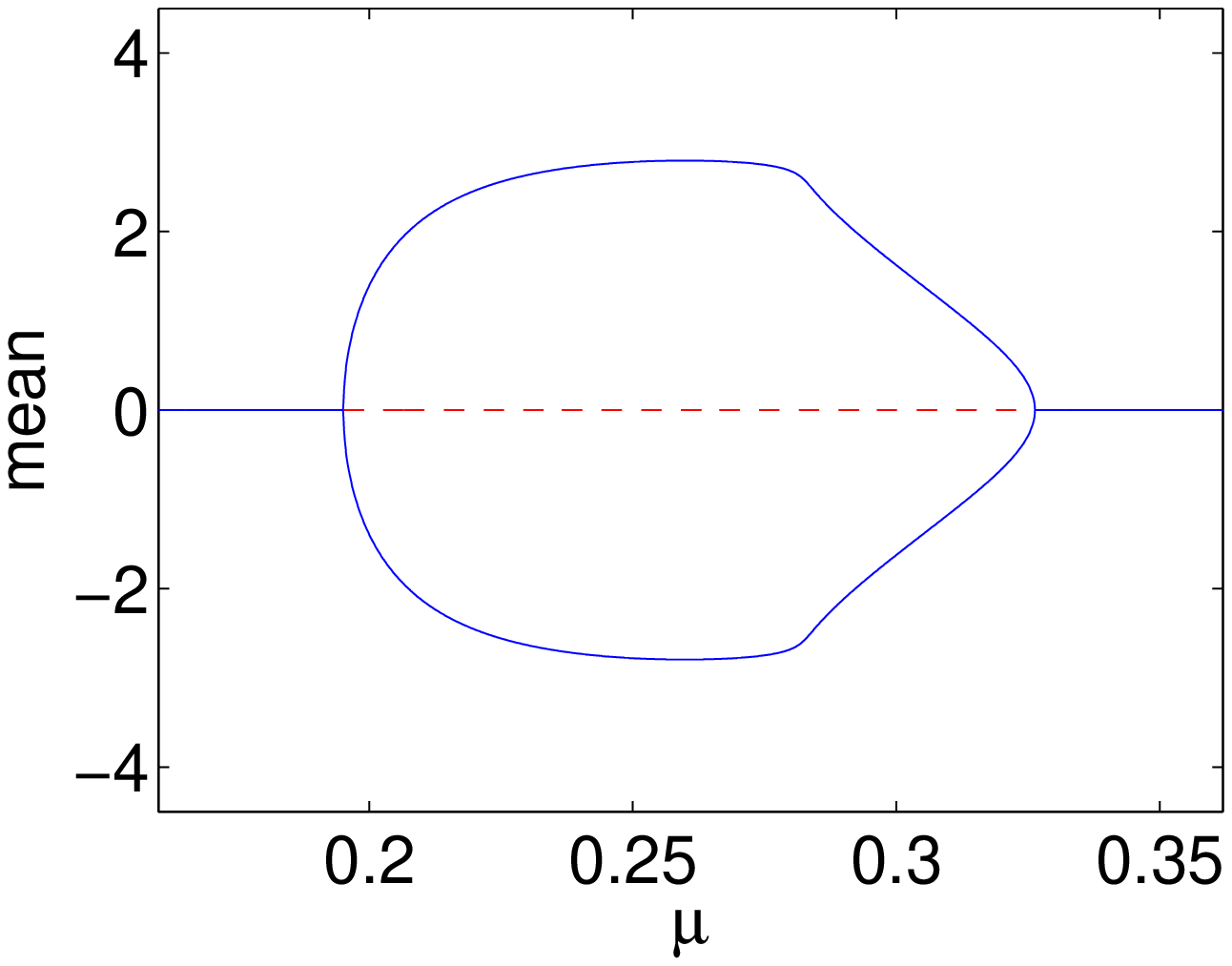}\newline
\caption{(Color online) The left panel: a segment of the top right panel in
Fig. \protect\ref{fig8}, for $g=-0.35$, clarifies details of the bifurcation
picture. The right panel: the asymmetry-measuring intergal characteristic, $%
\protect\int_{-\infty }^{+\infty }x|\protect\psi |^{2}dx$, for the
antisymmetric solution and the asymmetric one arising from it, as a function
of $\protect\mu $, in the same case as in the left panel. The loop
illustrating the asymmetric solution emerging from (through the
symmetry-breaking supercritical pitchfork) and merging back (through the
symmetry-restoring subcritical pitchfork) into the antisymmetric branch. The
notation is the same as in Fig. \protect\ref{fig3}. }
\label{fig9}
\end{figure}

As $g$ grows more negative, we notice that the symmetric and the
antisymmetric branches, while monotonically increasing with $\mu $
at small values of $N$, switch to become monotonically decreasing
functions of $\mu $ at a large value of $N$. Simultaneously, we
observe the bifurcation of two asymmetric states, each from a
different branch. While it is commonly known that attractive and
repulsive interactions favor bifurcations from symmetric and
anti-symmetric branches, respectively, here we encounter the first
(to our knowledge) example that features bifurcations from
\textit{both} branches. This unusual situation extends up to
$g=-0.32$. The complete bifurcation diagram is displayed in the top
right panel of Fig. \ref{fig8}, for the case of $g=-0.3$. In this
case, the symmetric and antisymmetric branches arise, as usual, from
their linear limits at $\mu =\omega _{0}$ and $\mu =\omega _{1}$,
respectively. A supercritical pitchfork bifurcation occurring (as
before) on the antisymmetric branch gives rise to an asymmetric
state at $\mu =-0.195$, destabilizing the antisymmetric one. What
is, however, different here is that this asymmetric state
``survives" only within a narrow parametric interval before it
\textit{merges back} into the antisymmetric branch through another
subcritical (``symmetry-restoring") pitchfork at $\mu =0.3263$. This
bifurcation loop is reminiscent of that reported previously for
two-component solitons in local 1D and 2D models with competing
self-focusing cubic and self-defocusing quintic nonlinearities
\cite{loop}. For higher $\mu $, another asymmetric
branch
arises from the \emph{symmetric} one through a {\it subcritical} 
pitchfork at $%
\mu =0.3465$; it is especially important to highlight that 
this pitchfork, which destabilizes the symmetric
branch, is a subcritical one, hence the asymmetric state is unstable too. It
is observed that, after its emergence, the norm of the asymmetric branch
decreases as $\mu $ decreases ($dN/d\mu >0$), before starting to rise at the
fold point at $\mu =0.2768$, featuring $dN/d\mu <0$ after that. The
stability of the state changes at the turning point, which is in agreement
with the well-known Vakhitov-Kolokolov (VK) criterion \cite{vk}, according
to which the slope of the branch determines its stability. In Fig. \ref{fig9}%
, a blowup provides a clearer view of these bifurcations.

The symmetric and the antisymmetric branches continue to increase their
norms with the growth of $\mu $, each of them featuring its own turning
point, at $\mu =0.391$ and $\mu =0.3853$ respectively, and change their
monotonicity thereafter. However, this change on the antisymmetric branch is
\textit{not} accompanied by a stability change. To demonstrate that this is
\emph{not} an exception to the VK criterion, we define the linearization
operators,
\begin{equation}
\begin{aligned} L_{+} &= L_{1} + L_{2} + 3 g |\psi|^{2},\\ L_{-} &= L_{1} -
L_{2} + g |\psi|^{2} ,\end{aligned}  \label{eqLpLm}
\end{equation}%
using notation $L_{1,2}$ defined in Eq. (\ref{eqL12}). It is known from Ref.
\cite{grill} that the instability arises, when the slope condition of the VK
criterion is violated, if $|n(L_{+})-n(L_{-})|=1$, $n$ standing for the
number of negative eigenvalues of each operator. The state under 
consideration is generically unstable if $|n(L_{+})-n(L_{-})|>1$. In our
case, the antisymmetric branch has $n(L_{-})\equiv 1$ (due to its
single-zero-crossing configuration profile, which is a zero mode of $L_{-}$).
Meanwhile, $n(L_{+})=1$ is true before the branch turns to the left, and
consequently $|n(L_{+})-n(L_{-})|=0$. Hence, even though the slope condition
is violated ($dN/d\mu >0$), the relevant theorem suggests that the VK
criterion does not actually apply here (and no stability change occurs).
After the turning point, the stability is expected since $n(L_{+})=2$, and
the slope condition is satisfied, which is in accordance with our observations.

Next, as the local interactions continue to grow stronger, another regime is
explored, as shown in the bottom left panel of Fig. \ref{fig8}, with $%
g=-0.35 $ as an example. In this case, the phenomenology of the symmetric
branch is similar to that in the previous situation; however, the two
sequential bifurcations taking place on the antisymmetric branch no longer
arise. Hence, the entire antisymmetric branch remains dynamically stable.
The asymmetric branch still emerges from the symmetric one through a
subcritical pitchfork, and later becomes stable past the fold point due to
the change of slope, in accordance with the VK criterion.

The above regime persists until $g=-0.38$. After that, the pitchfork
bifurcation occurring on the symmetric branch switches from subcritical to
supercritical, leading to the emergence of a stable asymmetric branch, as
shown in the last panel of Fig. \ref{fig8}. This shape of the bifurcation
diagram, consisting of the three branches, persists as far as $g$ decreases to $%
-1$. Note that during this process, the turning points of the symmetric and
antisymmetric branches keep getting lower (with respect to $N$), finally
leading to a situation where the turning points do not exist, and the two
branches immediately go left starting from their linear limits, i.e., the
symmetric and the antisymmetric solutions only exist for $\mu <\omega
_{0},\;\omega _{1}$ when $g<-0.69,\;-0.6$, respectively.

\section{Conclusions and future challenges}

In this work, we have presented 
a systematic 
study of the interplay between the
long-range nonlinear interactions (of either sign, repulsive or attractive)
and linear double-well potential (DWP) 
in the 1D setting. The two-mode 
approximation has been developed, that accounts for the nonlocality but
retains the structure similar to that established before in the case of the
contact interaction. This conclusion demonstrates that the fundamental
phenomenology of the spontaneous symmetry breaking (SSB) 
in the DWP persists
in the presence of the longer-range interactions, although the critical
properties themselves (e.g., the critical values of the chemical potential
and norm), at which the SSB bifurcation occurs, giving rise to the
asymmetric steady states, are sensitive to the precise interaction range. In
particular, our analysis has revealed a monotonous increase of the critical
values as a function of the interaction range, in the case of the
self-repulsion. A considerably more elaborate phenomenology was revealed by
the analysis in the context of competing long- and short-range interactions.
In that case, a delicate interplay between the strengths of the nonlocal
repulsion and local attraction gives rise, in addition to predominantly
attractive and predominantly repulsive regimes, to mixed ones with complex
bifurcation phenomena. On the one hand, symmetry-breaking effects were shown
to arise from each of the relevant solution branches; in some cases, they
are accompanied by reverse bifurcations, to form closed loops. On the other
hand, some of the previously studied supercritical bifurcations, such as the
one occurring on the symmetric branch, could become subcritical, being
subsequently coupled to additional fold bifurcations. All of these effects
are absent, to our knowledge, in models with a single cubic nonlinear term,
being consequences of the competition.

These results may be a motivation for studies in a number of future
directions. In particular, it would be relevant to consider the
generalizations of the DWP setting to higher dimensions, such as, the
four-well configuration, which was recently demonstrated to yield a much
richer phenomenology in the case of the contact interactions \cite{chenyu2d}. 
In the same connection, it is relevant to note that the few-mode approach,
used in this work for the consideration of the existence and stability of
the symmetric, antisymmetric, and asymmetric states, as functions of the
interaction strength, may be applied to other structures, such as
multi-dimensional bright \cite{Pedri05}, dark \cite{sann} and vortex \cite%
{Krolik} solitons, especially in cases where such structures are stabilized
by nonlocal nonlinearities. A natural objective of such an analysis may be
to unravel effects of the interaction range on structural stability of
nonlinear waveforms. Such studies are presently in progress and will be
reported elsewhere.

\acknowledgments PGK gratefully acknowledges support from NSF-DMS-0349023
(CAREER), NSF-DMS-0806762 and the Alexander-von-Humboldt Foundation. The
work of BAM was supported, in a part, by grant No. 149/2006 from the
German-Israel Foundation. The work of DJF was partially supported by the 
Special Account for Research Grants of the University of Athens.

\end{document}